\documentclass[11pt]{article}
\usepackage{enumitem}
\usepackage{natbib}
\usepackage{amsmath,amsthm,amssymb}
\usepackage{mathrsfs}
\usepackage{graphicx,psfrag,epsf}
\usepackage{epstopdf}
\usepackage{epsfig}
\usepackage{subcaption,multirow,booktabs}
\usepackage{bbm}	% for \mathbbm{1}
\usepackage{color}
\usepackage{mathtools} % coloneqq
\usepackage{authblk}
\usepackage{tikz}
\usetikzlibrary{arrows}
\usepackage{soul}
\usepackage{mathtools}
\usepackage{kotex}
\usepackage{comment}
\usepackage[toc,page]{appendix}

\usepackage{xr}
\RequirePackage[hyperindex,breaklinks,colorlinks,citecolor=blue,urlcolor=black]{hyperref} 

\setitemize{itemsep=-1mm}
\setenumerate{itemsep=-1mm}

\theoremstyle{plain}

\newtheorem{theorem}{Theorem}

\theoremstyle{definition}

\newtheorem{remark}{Remark}

\newcommand\bstheta{\boldsymbol{\theta}}

\newcommand{\bsSigma}{\boldsymbol\Sigma}
\newcommand{\tildeQ}{\tilde{\mathbf Q}}

\newcommand{\T}{\mathsf{T}}
\newcommand{\CD}{\mathrm{CD}}

\newcommand{\bsgamma}{\boldsymbol\gamma}

\newcommand{\blind}{0}

\makeatletter
\newcommand{\mylabel}[2]{#2\def\@currentlabel{#2}\label{#1}}
\makeatother

\begin{document}

\def\spacingset#1{\renewcommand{\baselinestretch}%
{#1}\small\normalsize} \spacingset{1}

%%%%%%%%%%%%%%%%%%%%%%%%%%%%%%%%%%%%%%%%%%%%%%%%%%%%%%%%%%%%%%%%%%%%%%%%%%%%%%

\if0\blind
{
      \title{\bf Bayesian Triangulation Splines: Spatial Adaptation on Irregular Domains
  }
\author[1]{Sihyeon Pyeon}
\author[2]{Sunwoo Lim}
%\author[1,3]{Jaewoo Park}
\author[1,3]{Seonghyun Jeong\thanks{ Corresponding author: \texttt{sjeong@yonsei.ac.kr}}}
\affil[1]{Department of Statistics and Data Science, Yonsei University}
\affil[2]{Marshall School of Business, University of Southern California}
\affil[3]{Department of Applied Statistics, Yonsei University}

  \maketitle
} \fi

\if1\blind
{
  \bigskip
  \bigskip
  \bigskip
  \begin{center}
    {\LARGE\bf Title}
\end{center}
  \medskip
} \fi

\begin{abstract}
Conventional nonparametric regression methods for two-dimensional non-rectangular domains often overlook domain geometry and allow smoothing across boundaries. In spatial and geostatistical applications, this assumption is frequently invalid because domain boundaries typically constrain interactions among observations. Accommodating spatially varying smoothness is also substantially more challenging than in the univariate setting, and most existing methods do not adequately capture this local structure of the target function.
To address these challenges, we propose Bayesian triangulation splines, which constructs locally adaptive splines over a polygonal domain. The method employs constrained Delaunay triangulations to respect boundary geometry and adapt to heterogeneous smoothness. A carefully designed prior further improves empirical performance. Under a global Sobolev smoothness assumption, we show that the proposed method achieves the optimal posterior contraction rate and adapts to unknown smoothness.
We also show that the method exhibits ideal spatial adaptation in the sense that it achieves the oracle rate for inhomogeneous or locally varying structural features. Crucially, this oracle guarantee is not specific to constrained Delaunay triangulations, but holds over any triangulation satisfying weak shape-regularity conditions. Simulation studies confirm that the proposed method outperforms existing approaches by achieving higher estimation accuracy while maintaining low model complexity.
\end{abstract}


\spacingset{1.0}

\section{Introduction}
Traditional spatial models often rely on the assumption of proximity-based smoothing according to Tobler’s first law of geography: ``Everything is related to everything else, but near things are more related than distant things.''
Although this principle is generally reasonable, it can fail in spatial and geostatistical applications on irregular domains, as domain boundaries often restrict interactions among observations. For example, ocean chlorophyll levels are typically smooth across open water but are interrupted by an isthmus, requiring smoothing methods that account for coastline boundaries \citep{wood2008soap}.
Similarly, sea salinity levels can be disrupted by physical barriers such as ice and landmasses in an archipelago, necessitating spatial modeling that properly reflects the complex boundaries \citep{jin2024spatial}.
A key aspect of modeling such data is properly accounting for spatial domain boundaries. Conventional approaches based on tensor product splines or Gaussian processes are not suitable in these settings. Ignoring domain boundaries can result in spurious smoothing artifacts by allowing information to flow across areas that should remain separated, such as holes and concavities.

%However, with the increasing interest in modeling geospatial data from complex domains—such as regions with irregular boundaries, concavities (e.g., bays and capes), or inner holes (e.g., lakes)—this assumption often fails in real-world applications. For example, \citet{wood2008soap} provided an instance of abrupt changes in the ocean chlorophyll levels caused by an isthmus that obstructs the water flow, though it is generally expected to vary smoothly within the water. Physical barriers, such as ice or landmasses, also induce discontinuities in sea salinity levels in adjacent regions. One can find a notable example in \citet{jin2024spatial}, which studied a distinctive pattern of the Arctic sea salinity caused by an archipelago. A key challenge in analyzing such data lies in accounting for domain constraints that disrupt spatial continuity. As noted by \citet{ramsay2002spline}, conventional methods based on tensor product splines assume that observations are obtained from a regular grid on a rectangular domain, which is not appropriate for applications involving constrained domains. When an interpolation is conducted ignoring the geometry of a region that consists of disjoint or well-separated subregions, the fitted function unintentionally introduces spurious relationships between these regions. This results in an artificial smoothing effect that connects areas that should remain independent, known as domain leakage. 

Several approaches, encompassing both frequentist and Bayesian perspectives, have been proposed to address this issue in the context of spatial smoothing.
\citet{ramsay2002spline} introduced finite element splines, extending smoothing splines to irregular domains by formulating the roughness penalty with differential operators restricted to the domain, implemented via a triangulation with natural boundary conditions.
\citet{wood2008soap} proposed the soap-film smoother (SFS), an analogy to physical membranes that offers greater boundary flexibility while remaining computationally straightforward and easily embedded in generalized additive models.
Building on spline approximation theory, subsequent work on bivariate penalized splines over triangulations (BPST) established a systematic framework using Bernstein-B\'ezier representations, providing stable bases, theoretical error bounds, and scalable algorithms for large data sets \citep{lai2013bpst,wang2020efficient,yu2020estimation}. On the Bayesian side,
\citet{niu2019intrinsic} developed a Gaussian process (GP) model for irregular domains by constructing a covariance structure from heat kernels, thereby preventing information from propagating across boundaries.
\citet{jin2024spatial} proposed BORA-GP, a scalable GP model that achieves computational efficiency through directed acyclic graph representations of dependence while accommodating domain geometry and barriers.
Based on a spanning tree construction, \citet{luo2021bast} introduced Bayesian additive spanning trees (BAST), an ensemble of weak learners designed to capture local variability over irregular domains.
Each of these methods offers distinct advantages and faces certain limitations. To assess their strengths and weaknesses, we consider two notions of adaptation: \emph{rate adaptation} and \emph{spatial adaptation}.

First, rate adaptation concerns the ability of an estimation procedure to attain the optimal convergence rate without knowing the smoothness of the target function. Convergence rates serve as a fundamental criterion in both frequentist and Bayesian inference. In frequentist analysis, they characterize the speed at which estimators converge to the true parameter. In Bayesian inference, posterior contraction rates describe the speed at which posterior distributions concentrate around the true parameter as the sample size increases. Among the related studies discussed above, convergence rate guarantees are available only for the frequentist triangulation-based approaches \citep{lai2013bpst,wang2020efficient,yu2020estimation}. The other frequentist methods and the Bayesian approaches reviewed above have not been accompanied by rate analysis. Although the frequentist triangulation-based estimators achieve minimax optimal rates under suitable conditions, they require prior knowledge of the smoothness parameter of the function class to balance estimation flexibility and model complexity. Therefore, none of the previously discussed methods achieves rate adaptation. In general, achieving rate adaptation is a long-standing and challenging problem in frequentist analysis \citep{lepskii1991problem,donoho1995adapting,birge1997model}. In contrast, Bayesian methods can often achieve rate adaptation more naturally by assigning a suitable prior distribution over model complexity \citep{belitser2003adaptive,van2009adaptive,arbel2013bayesian,shen2015adaptive}.

Second, spatial adaptation concerns the ability of a method to allocate model complexity according to local features of the target function. This property is especially important over irregular domains, where the target function may exhibit inhomogeneous smoothness while the boundary geometry restricts the relevant neighborhood structure. Spatial adaptation has been studied through theoretical formulations of local adaptivity \citep{donoho1994ideal,birge2001alternative}, but it is also often pursued in a practical sense as the ability to recover spatially varying smoothness.
Several frequentist approaches have been developed for this purpose, but they are often computationally demanding and algorithmically complex \citep{zhou2001spatially,miyata2003adaptive}. In contrast, Bayesian analogues are often more natural and can be implemented by placing a suitable prior over local model complexity \citep{smith1996nonparametric,denison1998automatic,dimatteo2001bayesian,chipman2010bart}. However, none of the related studies discussed above provides a theoretical guarantee for spatial adaptation over irregular domains.
From a practical perspective, BAST \citep{luo2021bast} is the only work designed to capture spatially varying smoothness among the methods addressing domain leakage. However, its convergence rate has not been established, and no oracle guarantee for spatial adaptation is available. In addition, we observe in our numerical studies that BAST often suffers from overfitting and is outperformed by competing methods.

\begin{figure}[t!]
\centering
\begin{subfigure}[t]{\textwidth}
\centering
\includegraphics[width=0.2\textwidth]{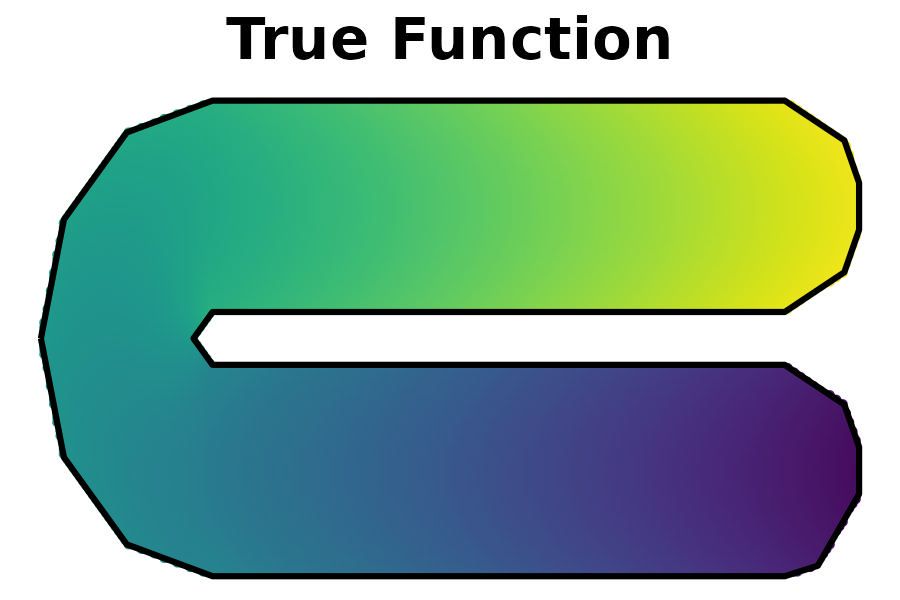} 
\includegraphics[width=0.2\textwidth]{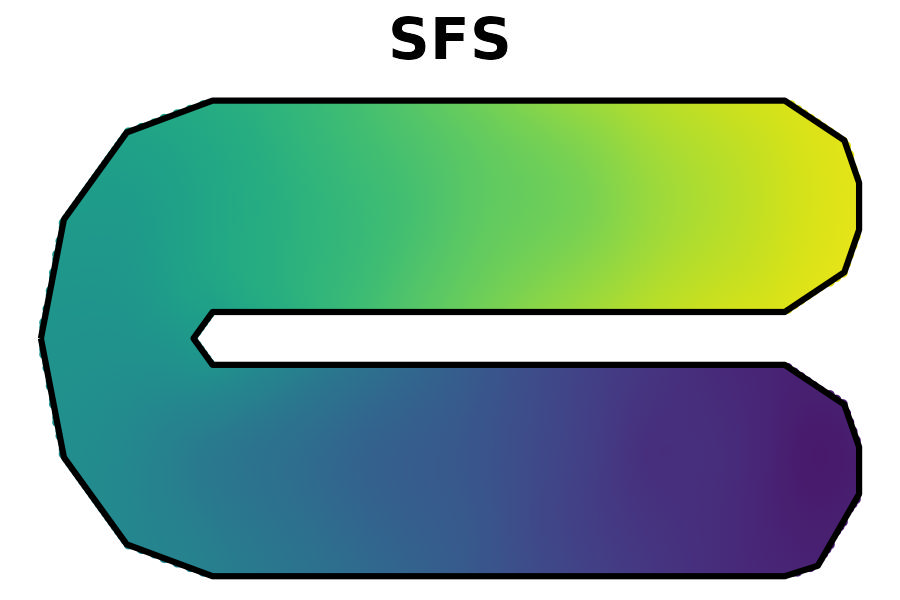} 
\includegraphics[width=0.2\textwidth]{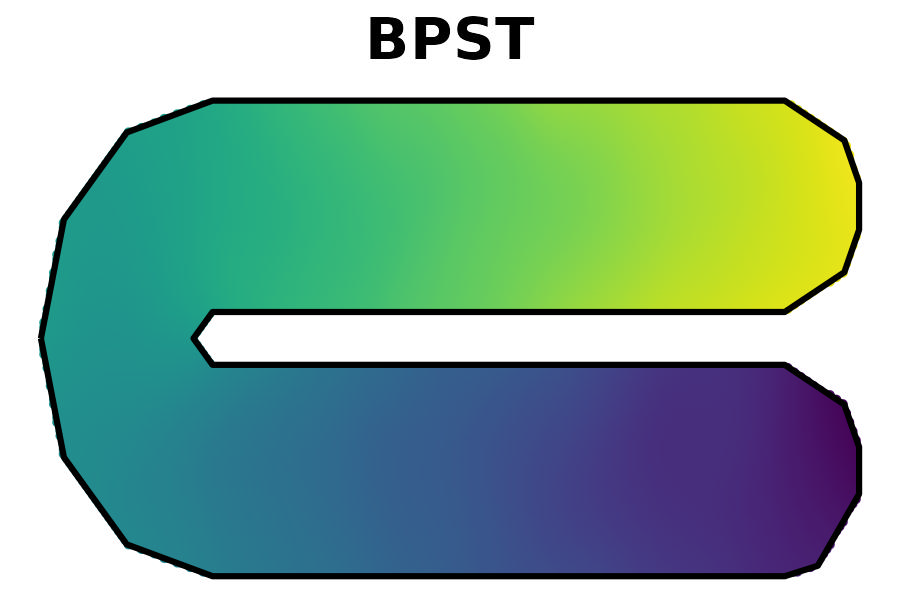} 
\\ \vspace{0.5em}
\includegraphics[width=0.2\textwidth]{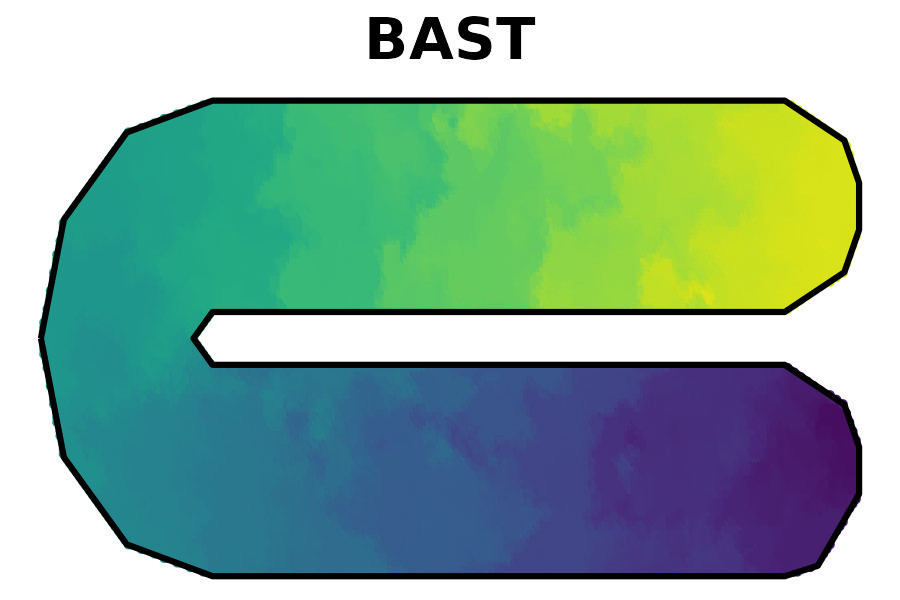} 
\includegraphics[width=0.2\textwidth]{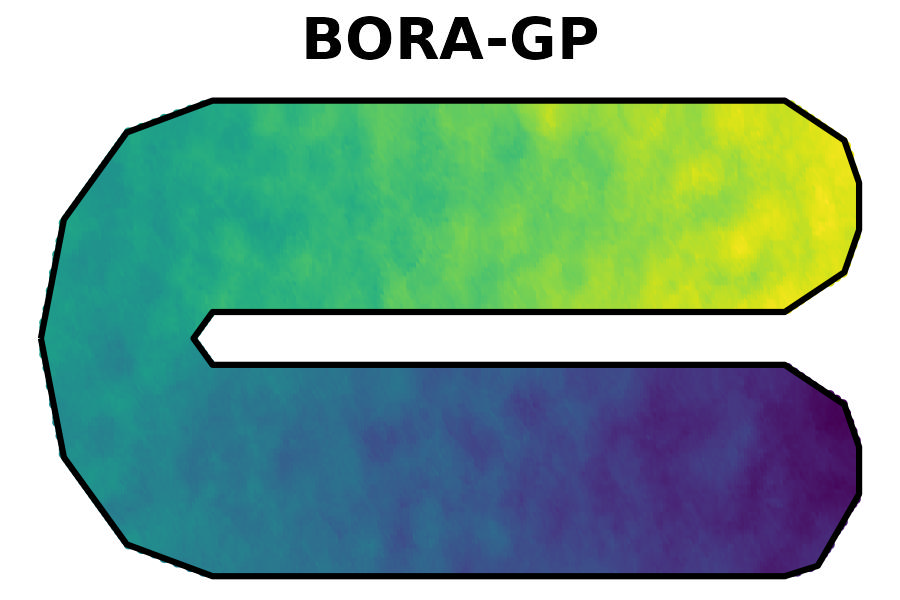} 
\includegraphics[width=0.2\textwidth]{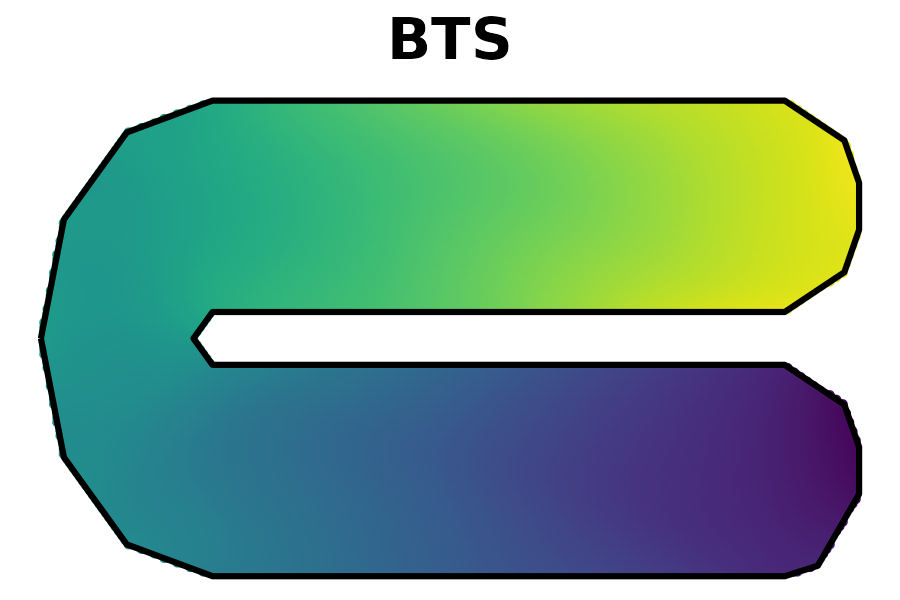}
\caption{Smooth target function with modest variation.}
\label{fig:compare_pred_ft}
\end{subfigure}

\vspace{1.5em}

\begin{subfigure}[t]{\textwidth}
\centering
\includegraphics[width=0.2\textwidth]{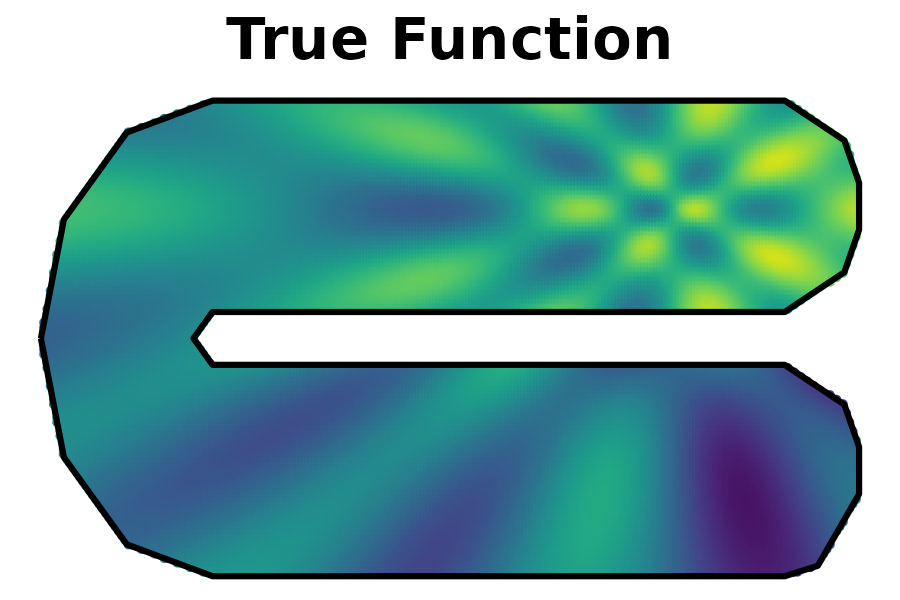} 
\includegraphics[width=0.2\textwidth]{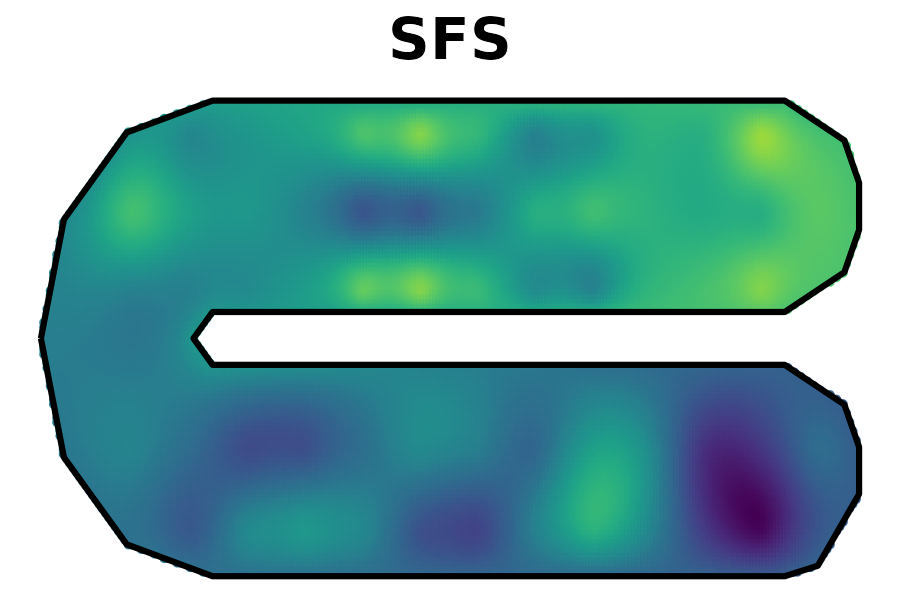} 
\includegraphics[width=0.2\textwidth]{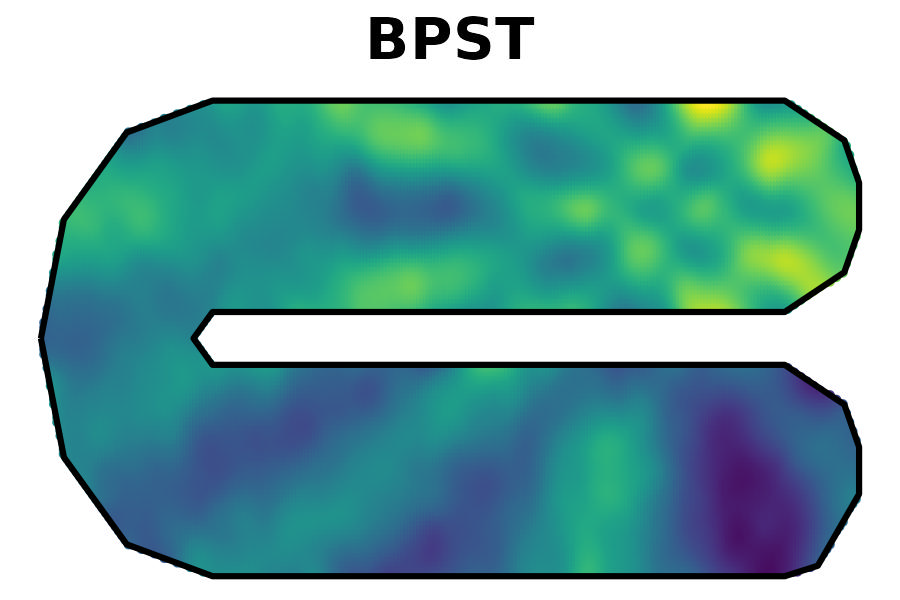} 
\\ \vspace{0.5em}
\includegraphics[width=0.2\textwidth]{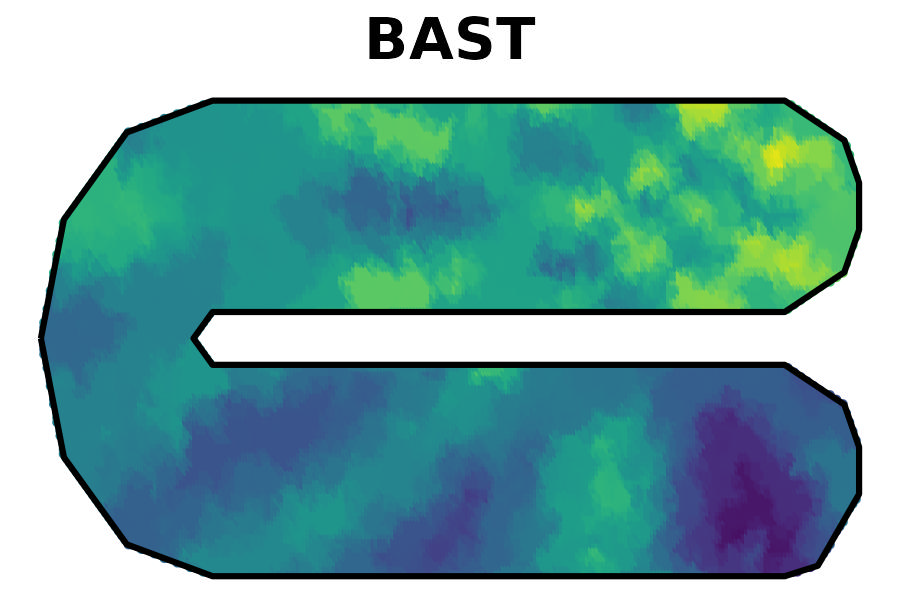} 
\includegraphics[width=0.2\textwidth]{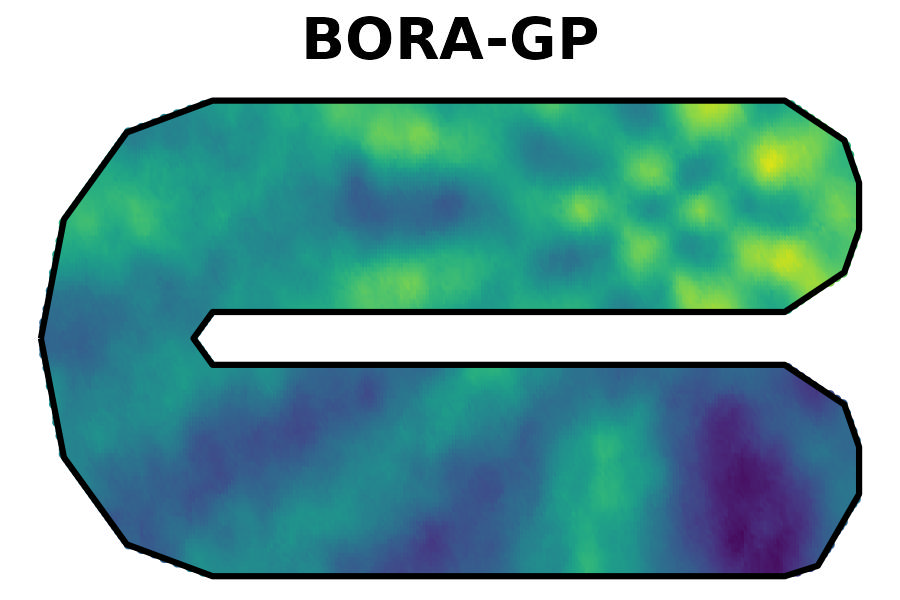} 
\includegraphics[width=0.2\textwidth]{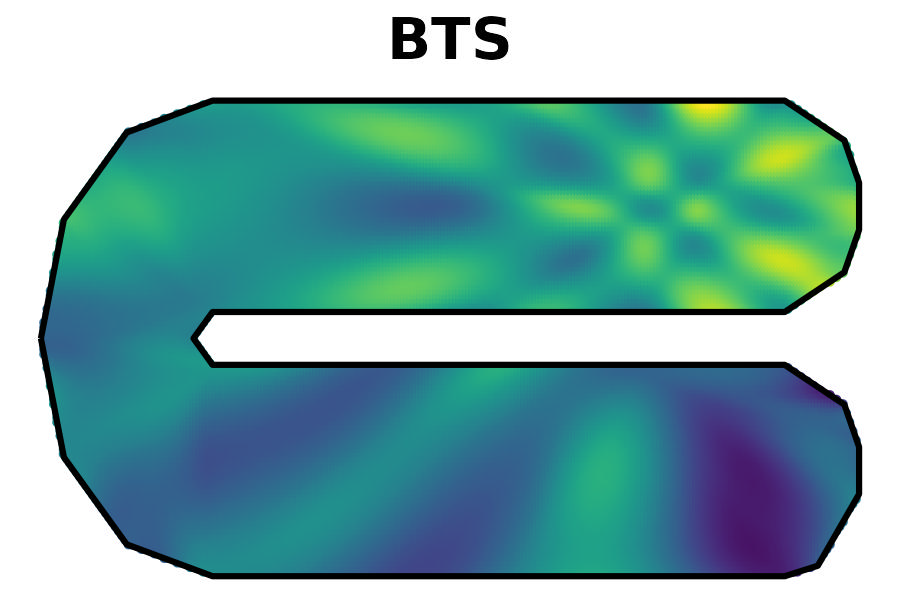}
\caption{Target function with spatially varying local complexity.}
\label{fig:compare_pred_ex}
\end{subfigure}

\caption{Pointwise mean predictions on a horseshoe domain. 
Each dataset consists of $n=5000$ training observations generated from the true function shown in the top-left panel of each subfigure with $\sigma = 0.5$.}
\label{fig:compare_pred}
\end{figure}

In this paper, we propose Bayesian triangulation splines (BTS) to address the limitations of existing methods for irregular domains. BTS builds on locally adaptive splines over an irregular domain constructed via \emph{constrained Delaunay triangulation} (CDT), thereby respecting complex boundary geometry. It achieves both forms of adaptation discussed above: near-minimax rate adaptation over global Sobolev classes and spatial adaptation through a near-oracle rate guarantee. Importantly, although the prior is constructed using CDT, the oracle benchmark is not confined to the CDT class; it ranges over the broader class of triangulations satisfying weak shape regularity. As discussed above, no existing method provides theoretical guarantees for these adaptation properties on irregular domains.
For posterior inference, we develop an efficient Markov chain Monte Carlo (MCMC) algorithm. The empirical advantage of BTS over competing methods is illustrated in Figure~\ref{fig:compare_pred}.

The rest of this paper is organized as follows. Section~\ref{sec:method} introduces the proposed methodology, along with a brief review of constrained Delaunay triangulations and bivariate splines on triangulations. Section~\ref{sec:prior} specifies the prior distributions, and Section~\ref{sec:mcmc} describes posterior inference. Section~\ref{sec:rate} presents theoretical results showing that the proposed procedure is rate adaptive in Sobolev spaces and achieves spatial adaptation. Section~\ref{sec:sim} provides simulation studies evaluating the performance of the proposed method. Section~\ref{sec:realdata} illustrates the method using a real dataset. Section~\ref{sec:disc} concludes the paper with some remarks.

\section{Bayesian Triangulation Splines}
\label{sec:method}

For each $i=1,\dots,n$, let $y_i\in\mathbb{R}$ denote the response variable and let $\mathbf{s}_i=(s_{i1},s_{i2})^\T\in\Omega$ represent the spatial location, where the domain $\Omega\subset\mathbb R^2$ is an open, bounded polygon. We consider the nonparametric regression model,
\begin{align}
y_i = f_0(\mathbf{s}_i) + \epsilon_i, \quad\epsilon_i\stackrel{\text{iid}}\sim N(0,\sigma_0^2),\quad i=1,\dots,n,
\label{eq:modelbig}
\end{align}
where $f_0:\Omega\rightarrow\mathbb{R}$ is an unknown bivariate regression surface and $\sigma_0^2>0$ is a variance parameter. Our aim is to characterize the properties of $f_0$ while respecting the geometry of $\Omega$.
Although we primarily focus on the modeling structure in \eqref{eq:modelbig}, additional terms can be incorporated into the mean response if needed. For example, if covariates $x_{ij}$, $i=1,\dots,n$, $j=1,\dots,p$, are available, one may include a linear predictor $\sum_{k=1}^p \beta_{0k}x_{ij}$ with coefficients $\beta_{0k}$, yielding $y_i = f_0(\mathbf{s}_i) +\sum_{k=1}^p \beta_{0k}x_{ij}+ \epsilon_i$, which corresponds to the form of partially linear models. Another extension is to add an additive component $\sum_{k=1}^p h_{0k}(x_{ij})$ with univariate functions $h_{0k}$, yielding an additive model $y_i = f_0(\mathbf{s}_i) +\sum_{k=1}^p h_{0k}(x_{ij})+ \epsilon_i$. Both extensions are straightforward to implement within the basis expansion framework.

The goal of this study is to develop an adaptive Bayesian procedure for modeling the bivariate function $f_0$. Specifically, we propose a method that adapts to spatial inhomogeneity while attaining the optimal posterior contraction rate without prior knowledge of the smoothness of $f_0$. At the same time, it must respect the complex boundary $\partial \Omega$ by preventing smoothing across it. To achieve these objectives, we employ locally adaptive splines on triangulations as described below.

\subsection{Bivariate Splines on Triangulation}
\label{sec:spline}

A triangulation $\Delta=\{\tau_1,\dots,\tau_{N}\}$ of a polygonal domain $\Omega$ is a finite collection of $N$ closed triangles $\tau_i$
that are pairwise disjoint except at common edges and vertices, and whose union equals the closure of $\Omega$; that is, for $i\ne j$, $\mu(\tau_i\cap\tau_j)=0$ and $\overline\Omega=\cup_{i=1}^{N} \tau_i$, where $\mu$ denotes the Lebesgue measure on $\mathbb R^2$.
Given $\Delta$, the spline space of degree $d$ and smoothness $r$ is defined as 
$$
\mathcal S_d^r(\Delta):= \{g\in \mathcal C^r(\Omega):g_{|\tau}\in\mathcal{P}_d(\tau),\tau\in\Delta\},
$$
where $\mathcal C^r(\Omega)$ denotes the space of functions $f:\Omega\rightarrow \mathbb R$ that are $r$-times continuously differentiable, and $\mathcal P_d(\tau)$ represents the space of polynomials of degree $d$ or less on a triangle $\tau\in\Delta$. 

A basis for $\mathcal S_d^r(\Delta)$ can be constructed using the Bernstein-B\'ezier representation.
For nonnegative integers $i,j,k$ with $i+j+k=d$, the Bernstein basis polynomials of degree $d$ relative to a triangle $\tau$ are defined as
\begin{align}
    B_{ijk}^{d,\tau} : \mathbf s \mapsto \frac{d!}{i!j!k!}b_1^i b_2^j b_3^k,\quad \mathbf s\in\tau,
    \label{eqn:bernstein}
\end{align}
where $(b_1,b_2,b_3)$ is the barycentric coordinate of $\mathbf s$ relative to $\tau$. The tuple $\mathcal B_\tau=(B_{ijk}^{d,\tau})_{i+j+k=d}:\tau\rightarrow\mathbb R^{(d+1)(d+2)/2}$ forms a basis for $\mathcal P_d(\tau)$, meaning any polynomial $ p \in \mathcal P_d(\tau)$ can be uniquely expressed as $p = \sum_{i+j+k=d}c_{ijk}B_{ijk}^{d,\tau}$ with coefficients $c_{ijk}$, which is known as the B-form of $p$ relative to $\tau$ (Theorem 2.4 of \citet{lai2007spline}). 
%For further properties of the B-forms, refer to Chapter 2 of \citet{lai2007spline}. 
%The collection $\mathcal B_\Delta = \cup_{\tau\in\Delta}\{\mathcal B_\tau\}$ then serves as a basis for the piecewise polynomial space over $\Delta$, which contains $\mathcal S_d^r(\Delta)$. 
We define the concatenated tuple $\mathcal B_\Delta=(B_{ijk}^{d,\tau})_{i+j+k=d,\tau\in\Delta}:\Omega\rightarrow \mathbb R^{|\mathcal B_\Delta|}$, where $|\mathcal B_\Delta|= N(d+1)(d+2)/2$. Then, every $f\in \mathcal S_d^r(\Delta)$ can be represented as $   f(\cdot) = \mathcal B_\Delta(\cdot)^\T\bsgamma_\Delta$ for some coefficient vector $\bsgamma_\Delta\in\mathbb R^{|\mathcal B_\Delta|}$.

However, $\mathcal B_\Delta$ does not form a basis for $\mathcal S_d^r(\Delta)$; rather, it spans an ambient space containing $\mathcal S_d^r(\Delta)$ without enforcing the $r$-smooth joins across triangle edges. To achieve a unique representation of $f\in \mathcal S_d^r(\Delta)$, additional linear constraints should be imposed on the spline coefficients. Specifically, $\bsgamma_\Delta$ must satisfy $\mathbf H_{\Delta}\bsgamma_\Delta=\mathbf 0$, where $\mathbf H_{\Delta}$ is a matrix encoding all the $r$-smooth linear constraints associated with the shared edges of $\Delta$. These constraints ensure the matching of directional derivatives of the B-forms across adjacent triangles up to order $r$ (Theorem~2.28 of \citet{lai2007spline}). 
Moreover, to facilitate a convenient prior specification described later, we explicitly separate the constant component from the spline representation.
Since $\mathcal B_\Delta(\cdot)^\T \mathbf 1_{|\mathcal B_\Delta|} =1
$ by the partition of unity, the coefficient vector in the direction of 
$\mathbf 1_{|\mathcal B_\Delta|}$ corresponds to a constant function. Therefore, to separate the intercept from the spline component, it is sufficient to impose the identifying constraint $\mathbf 1_{|\mathcal B_\Delta|}^\T \bsgamma_\Delta=0$. 

The two linear constraints are enforced by restricting $\bsgamma_\Delta$ to the null space of $[\mathbf H_{\Delta}^\T,\mathbf 1_{|\mathcal B_\Delta|}]^\T$, whose basis is obtained by the QR decomposition.
Toward this end, denote the QR decomposition of $[\mathbf H_\Delta^\T,\mathbf 1_{|\mathcal B_\Delta|}]$ by
$$
[\mathbf H_\Delta^\T,\mathbf 1_{|\mathcal B_\Delta|}] = [  \mathbf Q_\Delta, \tilde{\mathbf Q}_\Delta] \begin{bmatrix}
    \mathbf R_\Delta \\ \mathbf 0
\end{bmatrix},
$$ 
where $[  \mathbf Q_\Delta,\tilde{\mathbf Q}_\Delta ]$ is an orthogonal matrix and $ \mathbf R_\Delta$ is a full row rank upper triangular matrix.
%Since Lemma~? shows that $\mathbf H_{\Delta}\bsgamma_{\Delta}=\mathbf 0$ if and only if $ \bsgamma_\Delta = \tilde{\mathbf Q}_\Delta \boldsymbol\theta_\Delta$ for some $\boldsymbol\theta_\Delta$,
Since $\tilde{\mathbf Q}_\Delta$ forms a basis of the null space of $[\mathbf H_\Delta^\T,\mathbf 1_{|\mathcal B_\Delta|}]$,
we define $\tilde{\mathcal B}_\Delta=\tilde{\mathbf Q}_\Delta^\T \mathcal B_\Delta:\Omega\rightarrow \mathbb R^{J_\Delta}$, where $J_\Delta $ is the dimension of $\tilde{\mathcal B}_\Delta$. The spline space $\mathcal S_d^r(\Delta)$ is then expressed as 
\begin{align}
\label{eq:splinespace}
\mathcal S_d^r(\Delta)=\{g(\cdot)=\eta+\tilde{\mathcal B}_\Delta(\cdot)^\T\boldsymbol\theta_\Delta:\eta\in\mathbb R, \, \boldsymbol\theta_\Delta\in\mathbb R^{J_\Delta}\}.
\end{align}
That is, $(1,\tilde{\mathcal B}_\Delta)$ is a basis for $\mathcal S_d^r(\Delta)$.

\subsection{Constrained Delaunay Triangulation}
A triangulation $\Delta$ of $\Omega$ can be represented as a planar straight-line graph (PSLG) denoted by $(\mathcal V,\mathcal E)$, where $\mathcal V\subset \overline \Omega$ is the set of vertices and $\mathcal E\subset\{[\mathbf v,\mathbf v']:\mathbf v\ne \mathbf v'\in\mathcal V\}$ 
is the set of edges, where $[\mathbf v,\mathbf v'] = \{t\mathbf v +(1-t)\mathbf v':t\in[0,1]\}$ is the closed line segment with endpoints $\mathbf v$ and $\mathbf v'$. When a PSLG forms a triangulation, this representation induces the mapping $(\mathcal V,\mathcal E)\mapsto\Delta$,
emphasizing that to specify a triangulation $\Delta$, it suffices to determine its vertex and edge sets.
As discussed in Section~\ref{sec:spline}, bivariate splines on $\Delta$ depend on its underlying geometric structure: they are polynomials of degree $d$ on each triangle with $r$-smooth continuity across edges. To approximate $f_0$ effectively, it is therefore important to choose the vertex set $\mathcal V$ and the edge set $\mathcal E$ so as to avoid sliver triangles, namely triangles with extremely acute or obtuse angles. However, jointly controlling both $\mathcal V$ and $\mathcal E$ leads to an excessively large number of combinatorial possibilities. An effective strategy is to construct a triangulation solely from a given vertex set $\mathcal V$, with edges generated automatically according to geometric criteria that control triangle shape, while always including the boundary edges required to form $\partial\Omega$. In this way, the triangulation is completely determined by $\partial\Omega$ and $\mathcal V$. For many commonly used geometric criteria, such canonical triangulations are unique under mild genericity conditions, allowing the mapping to be simplified to $(\partial\Omega,\mathcal V)\mapsto\Delta$.

If $\Omega$ is convex, \textit{Delaunay triangulations} are particularly popular owing to their efficient construction \citep{delaunay1934sphere,cheng2013delaunay}. Delaunay triangulations subdivide the convex hull of a given point set into triangles whose circumcircles contain no other points.
This construction maximizes the smallest angle in the triangulation--a property known as the \textit{max-min angle property}--which helps to avoid sliver triangles. In our setting, however, Delaunay triangulations are not directly applicable because the domain $\Omega$ may be concave and strictly contained within the convex hull of the point set, and hence the domain boundary $\partial \Omega$ may not be properly respected.
To address this, we employ \textit{constrained Delaunay triangulations} (CDTs), which enforces prespecified edges as constraints \citep{chew1987constrained,cheng2013delaunay}.
%CDTs generalize Delaunay triangulations by allowing constrained edges. 
More precisely, a CDT requires that the circumcircle of each triangle contains no vertex that is visible from the triangle, where visibility means that the straight-line segment connecting the two points does not intersect the interior of any constrained edge.
Similar to Delaunay triangulations, CDTs satisfy the max-min angle property among triangulations with the same constrained edges \citep[Theorem 2.17]{cheng2013delaunay}. This property ensures that CDT constructions avoid sliver triangles, making them well suited for spline approximation on $\Omega$.

\begin{figure}
    \centering
    \includegraphics[width=0.5\linewidth]{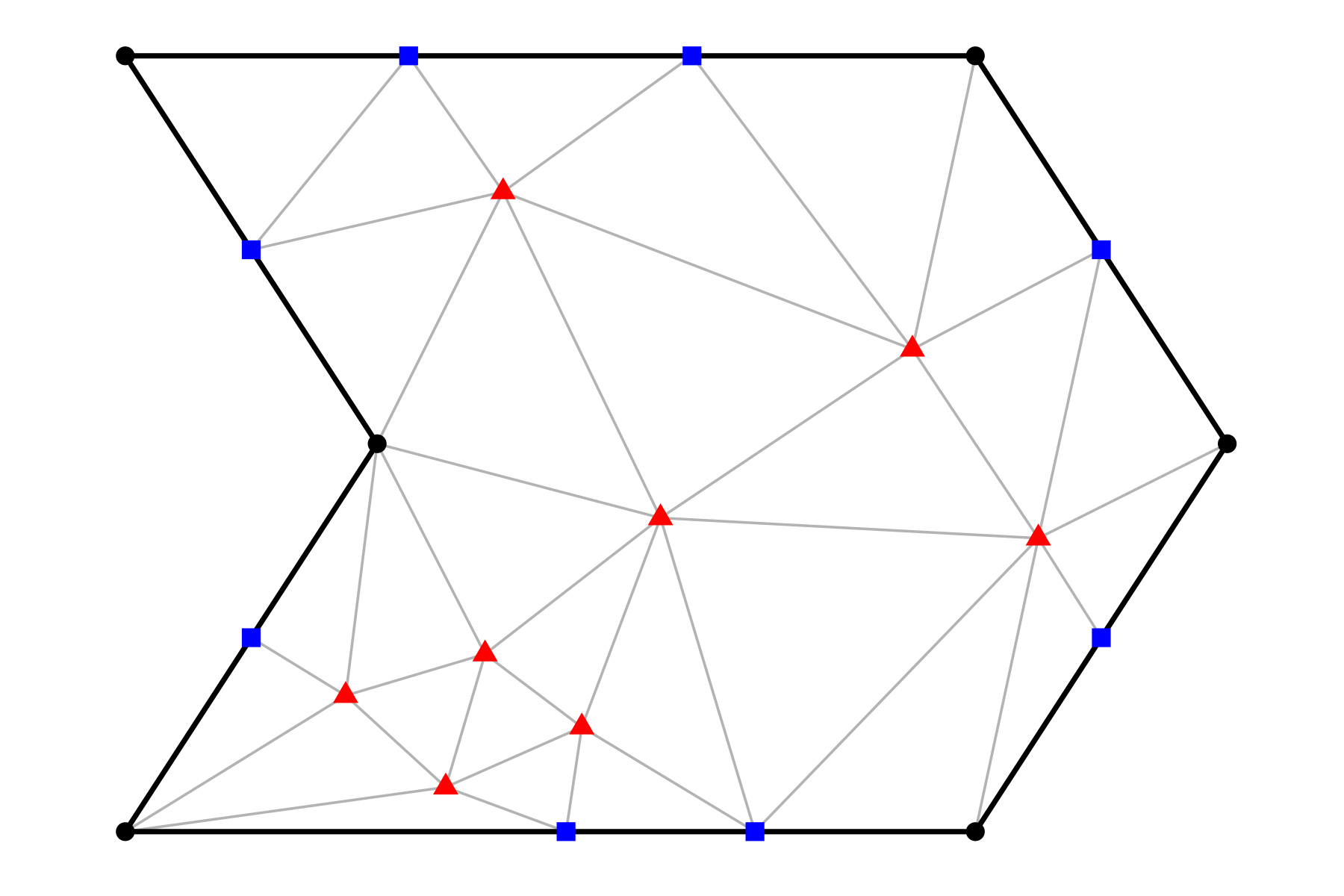}
    \caption{An example of a CDT of $\Omega$ with additional vertices. The corner vertices in $\mathcal V_\Omega$ are marked by black circles (\textcolor{black}{\large$\bullet$}), the boundary vertices in $\mathcal V_B$ by blue squares (\textcolor{blue}{$\scriptstyle\blacksquare$}), and the interior vertices in $\mathcal V_I$ by red triangles (\textcolor{red}{$\blacktriangle$}). The edges on $\partial \Omega$ are imposed as constrained edges in the construction of the CDT.}
    \label{fig:examplecdt}
\end{figure}

When constructing a CDT of $\Omega$, the constrained edges are chosen so as to preserve the domain boundary $\partial \Omega$.
To fully specify these constraints and complete the triangulation, additional boundary vertices may be placed along $\partial \Omega$, and interior vertices can be introduced within $\Omega$ as needed.
Specifically, let $\mathcal{V}_\Omega$ denote the vertex set of $\Omega$, and let $\mathcal{V}_B$ and $\mathcal{V}_I$ denote the additional vertices placed on $\partial\Omega$ and in $\Omega$, respectively. 
The constrained boundary edges are then generated from the vertex set $\mathcal V_\Omega \cup \mathcal{V}_B$ so as to coincide with $\partial \Omega$. The triangulation is completed by automatically generating the remaining edges to satisfy the visibility condition with respect to all vertices in $\mathcal V_\Omega \cup \mathcal V_B\cup\mathcal V_I$.
Several efficient algorithms for obtaining CDTs are available \citep{chew1987constrained,wang1987optimal}. In practice, CDTs can be easily constructed using convenient libraries. We use the Triangle library developed by Jonathan Shewchuk\footnote{\url{https://www.cs.cmu.edu/~quake/triangle.html}}, available through the R package \texttt{RTriangle}. Figure \ref{fig:examplecdt} shows an example of a CDT for a polygon $\Omega$.

It is worth noting that a CDT is unique for a given vertex set under a mild genericity condition. As in the case of Delaunay triangulations, if the vertices are in general position, meaning that no four or more points are cocircular, then the resulting CDT is unique \citep[Theorem~2.18]{cheng2013delaunay}. When multiple CDTs are possible for a given vertex set, we resolve the ambiguity by deterministically selecting one in lexicographic order, following the implementation in the Triangle library. This tie-breaking rule ensures that the CDT mapping  $(\partial\Omega,\mathcal{V}_\Omega,\mathcal{V}_B,\mathcal{V}_I)\mapsto \Delta$ is well defined, so that the triangulation is uniquely specified once the vertex set is fixed. Consequently, our Bayesian procedure only needs to determine the vertex sets $\mathcal{V}_B$ and $\mathcal{V}_I$ for the triangulation. In what follows, we denote by $\Delta_\Omega^\CD(\mathcal V_B,\mathcal V_I)$ the CDT of $\Omega$ generated by $\mathcal V_B$ and set $\mathcal V_I$. For $\zeta>0$ and $\ell>0$, let $\mathcal{T}_{\Omega}^{\CD}(\zeta,\ell)$ be the collection of all such CDTs whose minimum angle is at least $\zeta$ and whose minimum edge length is at least $\ell$.

\subsection{Spatially Adaptive Triangulation}
\label{sec:lat}

A CDT of $\Omega$ is determined by the vertex sets $(\mathcal V_B,\mathcal V_I)$. Thus, an appropriate choice of $\mathcal V_B$ and $\mathcal V_I$ is essential for obtaining a regular and adaptive triangulation. This choice should follow two main principles. First, the global complexity of the triangulation must balance approximation accuracy and model complexity. It should be large enough to provide an accurate approximation to $f_0$, but not so large as to impair computational efficiency or lead to overfitting. Second, the local placement of vertices determines the local resolution of the triangulation. Regions with smaller triangles provide greater local flexibility than regions with larger triangles, thereby allowing the spline estimator to adapt to spatial inhomogeneity.

Motivated by these principles, we treat both the number and locations of the vertices as unknown quantities and infer them from the data. This is naturally formulated in a Bayesian framework, where adaptivity can be induced through a prior distribution on the triangulation. A key ingredient is therefore a prior on $\mathcal V_B$ and $\mathcal V_I$ that controls the global complexity of the triangulation while allowing local refinement where the data support it. We specify this prior in the next section.

\section{Prior Specification}
\label{sec:prior}

\subsection{Prior for Triangulation}

The true regression surface $f_0$ is approximated within
$\mathcal S_d^r(\Delta)$. By the representation in
\eqref{eq:splinespace}, a prior on $\mathcal S_d^r(\Delta)$ is induced
through priors for $(\Delta,\eta,\bstheta_\Delta)$.
We first specify a prior over triangulations
$\Delta\in\mathcal T_\Omega^\CD(\zeta,\ell)$ with given $\zeta>0$ and $\ell>0$.
Since the CDT mapping
$(\partial\Omega,\mathcal V_\Omega,\mathcal V_B,\mathcal V_I)
\mapsto \Delta_\Omega^\CD(\mathcal V_B,\mathcal V_I)$
is assumed to be well defined, it suffices to assign a prior to
$(\mathcal V_B,\mathcal V_I)$.

Let $
V_B=|\mathcal V_B|$ and $V_I=|\mathcal V_I|$.
Let $\mu_{\partial\Omega}$ denote arclength measure on $\partial\Omega$
and let $\mu_\Omega$ denote Lebesgue measure on $\Omega$. We define the
dominating measure
$
\mu_\Omega^\oplus
=
\sum_{v_B=0}^\infty\sum_{v_I=0}^\infty
\mu_{\partial\Omega}^{v_B}\otimes\mu_\Omega^{v_I} 
$, on the disjoint union $\bigsqcup_{v_B,v_I} (\partial\Omega)^{v_B}\times \Omega^{v_I}$,
where the zeroth product measure is interpreted as the unit mass on the
empty configuration.
Define the admissible set of vertices,
$$
\mathcal A(\zeta,\ell)
=
\left\{(\mathcal V_B,\mathcal V_I):
\Delta_\Omega^{\CD}(\mathcal V_B,\mathcal V_I)\in\mathcal T_\Omega^\CD(\zeta,\ell)
\right\}.
$$
We assign a prior to $(\mathcal V_B,\mathcal V_I)$ by specifying its
density with respect to $\mu_\Omega^\oplus$ as
\begin{align}
\frac{d\Pi_{\mathcal V}}{d\mu_\Omega^\oplus}
(\mathcal V_B,\mathcal V_I)
\propto
\frac{
a_B(V_B)a_I(V_I)
\mathbbm 1_{\mathcal A(\zeta,\ell)}(\mathcal V_B,\mathcal V_I)
}{
|\partial\Omega|^{V_B}|\Omega|^{V_I}
},
\label{eq:prior-vtx-density}
\end{align}
where $a_B(V_B) = e^{-C_B V_B\log V_B}$ and $a_I(V_I) = e^{-C_I V_I\log V_I}$ with prespecified constants $C_B>0$ and $C_I>0$.
The omitted normalizing constant in \eqref{eq:prior-vtx-density} is independent of $(V_B,V_I)$ and cancels from all
Metropolis-Hastings ratios. %Hence no fixed-count admissible-volume constant needs to be evaluated.
The factors $a_B$ and $a_I$ 
serve as a complexity penalty on the numbers of additional boundary and
interior vertices. The induced marginal prior on $(V_B,V_I)$ is tilted by the
admissible configuration volume.
The induced prior distribution over
$\mathcal T_\Omega^{\CD}(\zeta,\ell)$ is then given by the pushforward
measure
\begin{align*}
\Pi_\Delta
=
\Pi_{\mathcal V}\circ(\Delta_\Omega^{\CD})^{-1}.
\end{align*}

\subsection{Prior for Spline Coefficients}

Recall that the spline space is characterized by the basis representation in \eqref{eq:splinespace}.
To complete the prior specification on $\mathcal S_d^r(\Delta)$, we place a prior on the coefficients $(\eta,\boldsymbol\theta_\Delta)$ conditional on the triangulation $\Delta$.
The prior for $(\eta,\boldsymbol\theta_\Delta)$ must be chosen carefully to reflect the model selection nature of the proposed framework \citep{moreno1998intrinsic}.
The intercept $\eta$ is separated from the triangulation search and can be assigned a weakly informative prior.
In contrast, a diffuse prior on $\boldsymbol\theta_\Delta$ is unsuitable because it may lead to Bartlett's paradox \citep{bartlett1957comment}.
We therefore construct a prior for $\boldsymbol\theta_\Delta$ by combining a baseline coefficient-size penalty with a roughness penalty induced by the B-form representation.

Recall that under the linear constraints
$\mathbf H_\Delta \bsgamma_\Delta=0$ and $\mathbf 1_{|\mathcal B_\Delta|}^\T \bsgamma_\Delta=0$, the restricted spline $\mathcal B_\Delta(\cdot)^\T\bsgamma_\Delta$ represents a function in $\mathcal S_d^r(\Delta)$ with the constant component removed.
Since $\mathcal B_\Delta$ forms a partition of unity and hence has a comparable scale across its elements,
a prior induced by
the ridge penalty $\bsgamma_\Delta^\T \bsgamma_\Delta$ is reasonable to control the size of the spline $\mathcal B_\Delta(\cdot)^\T\bsgamma_\Delta$.
Since $\bsgamma_\Delta = \tilde{\mathbf Q}_\Delta \boldsymbol\theta_\Delta$ and $\tilde{\mathbf Q}_\Delta^\T \tilde{\mathbf Q}_\Delta = \mathbf I_{J_\Delta}$, this leads to a Gaussian prior for $\bstheta_\Delta$ with the quadratic term $\bstheta_\Delta^\T \bstheta_\Delta$.
However, a coefficient-size penalty alone does not explicitly control the roughness of the fitted surface.
Motivated by \citet{lim2023synergizing}, who observed that incorporating roughness penalization into a model selection prior can reduce the modeling bias induced by finite truncation, we augment the baseline ridge penalty with a roughness penalty.
Following the triangulation literature \citep{lai2013bpst,yu2020estimation}, we employ the thin plate spline penalty of a function $f:\Omega\rightarrow\mathbb R$, given by
\begin{align}
    \mathcal E(f) = \int_\Omega \left\{(\nabla_{s_1}^2f)^2+2(\nabla_{s_1}\nabla_{s_2}f)^2+(\nabla_{s_2}^2 f)^2\right\}d{s_1}d{s_2}, 
   \label{eq:penalty}
\end{align}
where $\nabla_{s_j}^qf$ denotes the $q$-th order derivative of $f$ in the direction of $s_j$ for $j=1,2$.
The penalty applied to the spline $\mathcal B_\Delta(\cdot)^\T\bsgamma_\Delta$ can be expressed as
\begin{align*}
  \mathcal E({\mathcal B}_\Delta^\T\bsgamma_\Delta) = \bsgamma_\Delta^\T \mathbf P_\Delta\bsgamma_\Delta = \boldsymbol\theta_\Delta^\T \tilde{\mathbf Q}_\Delta^\T\mathbf P_\Delta
\tilde{\mathbf Q}_\Delta\boldsymbol\theta_\Delta,
\end{align*}
for some $\mathbf P_\Delta\in\mathbb R^{|\mathcal B_\Delta|\times |\mathcal B_\Delta|}$ a block diagonal matrix whose blocks impose the penalty in \eqref{eq:penalty} on the B-form over each $\tau\in\Delta$ (see the supplementary material for more details).
By combining the two terms as a convex combination in the precision matrix, the proposed prior for $(\eta,\boldsymbol\theta_\Delta)$ is given by
\begin{align} 
\begin{split}
        \eta\mid\sigma^2 &\sim \text{N}(0,\kappa^2\sigma^2),  \\
\boldsymbol\theta_\Delta \mid \Delta,\sigma^2,\nu,\lambda &\sim \text{N}_{J_\Delta}\!\left(\mathbf 0_{J_\Delta},\lambda\sigma^2\!\left(\frac{\nu}{V_\Delta} \tilde{\mathbf Q}_\Delta^\T\mathbf P_\Delta\tilde{\mathbf Q}_\Delta+(1-\nu)\mathbf I_{J_\Delta}\right)^{-1}\right), 
\end{split}
\label{eq:priortheta}
\end{align}
where $\kappa^2>0$ is a sufficiently large constant, $V_\Delta=|\mathcal V_\Omega|+|\mathcal V_B|+|\mathcal V_I|$ denotes the total number of vertices of the triangulation $\Delta$, $\lambda>0$ is a dispersion parameter, and $\nu\in(0,1)$ is a weight parameter that balances the two penalties. The factor $V_\Delta$ is introduced to appropriately scale the eigenvalues of $\tilde{\mathbf Q}_\Delta^\T\mathbf P_\Delta\tilde{\mathbf Q}_\Delta$ appropriately for the theoretical development, but it has little empirical effect. 
As $\nu\rightarrow 1$, the prior consists only of the term roughness penalty term on $\boldsymbol\theta_\Delta$. Conversely, as $\nu\rightarrow 0$, the prior reduces to ridge penalization on $\boldsymbol\theta_\Delta$.

To complete the prior specification, we assign priors with exponential tails on $\nu$ and $\lambda$, ensuring that for some $k>0$ and any increasing sequence $a_n>0$,
\begin{align}
\begin{split}
        \log \Pi\{\nu > 1-a_n^{-k}\}&\lesssim -a_n, \\
       \log \Pi\{\lambda > a_n^k\} &\lesssim -a_n.
\end{split}
\label{eq:priortaulambda}
\end{align}
These tail properties are essential to satisfy the theoretical requirements.
Note that an inverse-gamma prior for $\lambda$, despite its semi-conjugacy, does not satisfy \eqref{eq:priortaulambda} because of its polynomial right tail. Among priors with exponential tails, we adopt an exponential prior with rate $c_\lambda>0$ for $\lambda$. This choice yields a generalized inverse Gaussian conditional posterior for $\lambda$, thereby allowing convenient posterior updates.
For $\nu$, computation can be efficiently performed using grid sampling for any prior distribution. To satisfy \eqref{eq:priortaulambda}, we use a uniform prior on $(0, 1 - \delta_\nu)$ with a small constant $\delta_\nu > 0$.
Lastly, for the variance parameter $\sigma^2$, we impose an inverse gamma prior, \begin{align}
\sigma^2\sim \text{IG}(a_\sigma,b_\sigma),
\label{eq:IGprior}
\end{align}
with small $a_\sigma>0$ and $b_\sigma>0$.  This specification is a natural choice because of its conjugacy, allowing complete marginalization in the posterior and thereby enabling efficient posterior updates for triangulations.
Some additional technical difficulties arise from the polynomial tails of the induced marginal $t$-prior on $(\eta, \bstheta_\Delta)$. We address this issue by directly analyzing the marginal posterior of $\sigma^2$ and truncating regions with negligible posterior mass; see the supplementary material.

\begin{remark}
As discussed in Section~\ref{sec:method}, the mean response may include additional components beyond $f_0$. Priors for these components can be incorporated straightforwardly given their parameterization. For example, if a linear predictor with $p$-dimensional parameters is included, a Gaussian prior is a natural choice, with additional shrinkage potentially beneficial when $p$ is large. If a nonparametric additive component is included, standard constructions such as spline-based representations with coefficient priors or Gaussian process priors can be employed. As these extensions follow directly from standard formulations, we omit further details.
\end{remark}

\section{Posterior Inference via Markov chain Monte Carlo}
\label{sec:mcmc}

This section elaborates the MCMC algorithm that explores the joint posterior distribution $\pi(\eta, \bstheta_\Delta, \Delta, \sigma^2, \nu, \lambda\mid\mathbf y)$. After marginalizing out some parameters, the triangulation can be updated using an appropriately designed proposal rule. Specifically, we use birth, death, and move proposals for updating the triangulation. 
The remaining parameters are straightly updated from their conditional distributions.

\subsection{Sampling Steps}
\label{sec:sampling}

We first present the overall sampling steps. The priors introduced in Section~\ref{sec:prior} are carefully chosen to facilitate the construction of the sampler. As described, to meet the tail conditions in \eqref{eq:priortaulambda}, we assign an exponential prior and a truncated uniform prior on $\lambda$ and $\nu$, respectively. Although an exponential prior for $\lambda$ is not conjugate, its monotone density enables a straightforward sampling scheme via data augmentation. For $\nu$, we employ grid sampling with efficiently evaluated density values at the specified grid points.
We define $\mathbf W_\Delta = [\mathbf 1_n, \tilde{\mathbf B}_\Delta]\in \mathbb R^{n\times (J_\Delta+1)}$, where $\tilde{\mathbf B}_\Delta \in \mathbb R^{n\times J_\Delta}$ is the basis matrix with its $i$th row given by $\tilde{\mathcal B}_\Delta(\mathbf s_i)$. We also define the block diagonal matrix $\bsSigma_{\Delta,\nu,\lambda}=\mathrm{diag}(\kappa^2,\lambda(\nu V_\Delta^{-1}\tilde{\mathbf Q}_\Delta^\T \mathbf P_\Delta \tilde{\mathbf Q}_\Delta+(1-\nu)\mathbf I_{J_\Delta})^{-1})$. The following describes a blocked Gibbs sampler for exploring the posterior distribution.

\begin{enumerate}[label=(\roman*)]
    \item Draw $\Delta$ from $\pi(\Delta \mid \mathbf{y}, \nu, \lambda) \propto \pi(\Delta) p(\mathbf{y} \mid \Delta, \nu, \lambda)$ using the birth-death-move proposals described in Section~\ref{sec:bdm}. The marginal likelihood $p(\mathbf{y} \mid \Delta, \nu, \lambda)$ is given by
\begin{align*}
    \textstyle p(\mathbf{y} \mid \Delta, \nu, \lambda) &\propto  
    \Big\lvert \mathbf I_{J_\Delta+1} - 
    \left( \mathbf W_\Delta^\T\mathbf W_\Delta + \bsSigma_{\Delta,\nu,\lambda}^{-1} \right)^{-1} 
    \mathbf W_\Delta^\T \mathbf W_\Delta\Big\rvert^{1/2} \\
    &\quad \times \left[ b_\sigma + \frac{1}{2} \Big( \mathbf y^\T \mathbf y - \mathbf y^\T \mathbf W_\Delta  
    \left( \mathbf W_\Delta^\T \mathbf W_\Delta + \bsSigma_{\Delta,\nu,\lambda}^{-1} \right)^{-1} 
    \mathbf W_\Delta^\T \mathbf y \Big) \right]^{-(a_\sigma + n/2)}.
\end{align*}
The details of the Metropolis update are given in Section~?.

    \item Draw $\sigma^2$ from $\pi(\sigma^2 \mid \mathbf{y}, \Delta, \nu, \lambda)$, which is 
\begin{align*}
    \text{IG}\!\left( a_\sigma + \frac{n}{2}, b_\sigma + \frac{1}{2}\left(\mathbf y^\T \mathbf y - \mathbf y^\T \mathbf W_\Delta \left( \mathbf W_\Delta^\T \mathbf W_\Delta + \bsSigma_{\Delta,\nu,\lambda}^{-1} \right)^{-1}\mathbf W_\Delta^\T \mathbf y \right) \right).
\end{align*}

    \item Draw $(\eta, \bstheta_\Delta)$ from $\pi(\eta, \bstheta_\Delta \mid \mathbf{y}, \Delta, \nu, \lambda, \sigma^2) $, which is
\begin{align*}
    \text{N}_{J_\Delta+1}\!\left(  \left(\mathbf W_\Delta^\T \mathbf W_\Delta + \bsSigma_{\Delta,\nu,\lambda}^{-1} \right)^{-1}\mathbf W_\Delta^\T \mathbf y, \sigma^2 \!\left(\mathbf W_\Delta^\T \mathbf W_\Delta + \bsSigma_{\Delta,\nu,\lambda}^{-1} \right)^{-1} \right).
\end{align*}

    \item Draw $\nu$ from $\pi(\nu \mid \mathbf{y}, \Delta, \lambda, \sigma^2 , \eta, \bstheta_\Delta) \propto \pi(\nu) \pi(\bstheta_\Delta \mid \Delta,\nu, \lambda, \sigma^2)$ using grid sampling. Specifically,
\begin{align*}
\pi(\bstheta_\Delta\mid\Delta,\nu,\lambda,\sigma^2) &\propto \nu^{{J_\Delta}/{2}} \prod_{k=1}^{J_\Delta}\left(\frac{\rho_k(\tildeQ_\Delta^\T\mathbf P_\Delta\tildeQ_\Delta)}{V_\Delta} + \frac{1-\nu}{\nu}\right)^{1/2}\\ &\quad\times\exp\left(-\frac{\nu}{2\lambda\sigma^2V_\Delta}\bstheta_\Delta^\T\tildeQ_\Delta^\T\mathbf P_\Delta \tildeQ_\Delta\bstheta_\Delta-\frac{1-\nu}{2\lambda\sigma^2}\bstheta_\Delta^\T\bstheta_\Delta\right),
\end{align*}
where $\rho_k$ denotes the $k$th eigenvalue of the matrix. This sampling construction is highly efficient, as time-consuming operations such as eigen-decomposition and matrix multiplication need to be computed only once during the density evaluation. %See {\color{red} Section~?} for further details.

    \item Draw $\lambda$  from $\pi(\lambda \mid \mathbf{y}, \Delta, \nu, \sigma^2 , \eta, \bstheta_\Delta)$, which is
    \begin{align*}        
        \lambda \mid \mathbf{y}, \Delta, \nu, \sigma^2 , \eta, \bstheta_\Delta &\sim \text{GIG}\!\left(2c_\lambda,\frac{\nu}{\sigma^2V_\Delta}\bstheta_\Delta^\T\tildeQ_\Delta^\T\mathbf P_\Delta \tildeQ_\Delta\bstheta_\Delta+\frac{1-\nu}{\sigma^2}\bstheta_\Delta^\T\bstheta_\Delta, \frac{2-J_\Delta}{2} \right),
    \end{align*}
where $\mathrm{GIG}(a,b,p)$ denotes the generalized inverse Gaussian distribution with density proportional to $t \mapsto t^{p-1}\exp{-(a t + b/t)/2}$ for $t>0$, where $a>0$, $b>0$, and $p\in\mathbb{R}$.
\end{enumerate}

\subsection{Birth-Death-Move Proposals for Triangulation}
\label{sec:bdm}

We first specify a proposal rule for updating triangulations over
$\mathcal T_\Omega^\CD(\zeta,\ell)$. Since the CDT $\Delta_\Omega^\CD(\mathcal V_B,\mathcal V_I)$ is uniquely determined by
the vertex sets $(\mathcal V_B,\mathcal V_I)$, updating
$\Delta$ reduces to updating the vertices in $\mathcal V_B$ and  $\mathcal V_I$. The corner vertices in $\mathcal V_\Omega$
are kept fixed throughout the Markov chain.

\begin{figure}[t!]
    \centering
    \begin{minipage}[c]{0.3\textwidth}
        \centering
        \begin{subfigure}[b]{\textwidth}
            \centering
            \includegraphics[width=\textwidth]{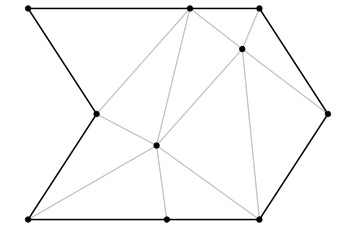}
            \subcaption{Initial triangulation}
            \label{fig:initial}
        \end{subfigure}
    \end{minipage}
    \hfill
    \begin{minipage}[c]{0.68\textwidth}
        \centering
        % --- 첫 번째 줄: Interior Proposals ---
        \begin{subfigure}[b]{0.31\textwidth}
            \centering
            \includegraphics[width=\textwidth]{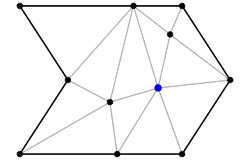}
            \subcaption{Interior birth}
            \label{fig:int_birth}
        \end{subfigure}
        \hfill
        \begin{subfigure}[b]{0.31\textwidth}
            \centering
            \includegraphics[width=\textwidth]{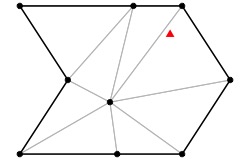}
            \subcaption{Interior death}
            \label{fig:int_death}
        \end{subfigure}
        \hfill
        \begin{subfigure}[b]{0.31\textwidth}
            \centering
            \includegraphics[width=\textwidth]{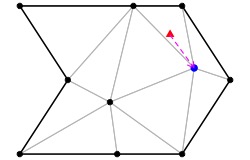}
            \subcaption{Interior move}
            \label{fig:int_move}
        \end{subfigure}
        
        \vspace{2ex} 
        
        % --- 두 번째 줄: Boundary Proposals ---
        \begin{subfigure}[b]{0.31\textwidth}
            \centering
            \includegraphics[width=\textwidth]{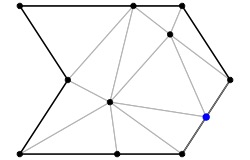} 
            \subcaption{Boundary birth}
            \label{fig:bnd_birth}
        \end{subfigure}
        \hfill
        \begin{subfigure}[b]{0.31\textwidth}
            \centering
            \includegraphics[width=\textwidth]{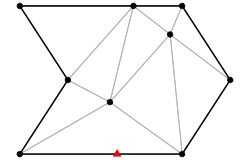} 
            \subcaption{Boundary death}
            \label{fig:bnd_death}
        \end{subfigure}
        \hfill
        \begin{subfigure}[b]{0.31\textwidth}
            \centering
            \includegraphics[width=\textwidth]{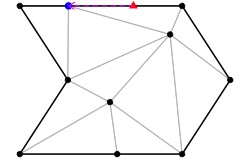}
            \subcaption{Boundary move}
            \label{fig:bnd_move}
        \end{subfigure}
    \end{minipage}
    
    % 전체 캡션 고쳐서 사용 가능
    \caption{Examples of birth-death-move proposals.}
    \label{fig:BDM_extended}
\end{figure}

At each iteration, we choose one of the six proposal families
\[
\mathcal M
=
\{B^+, I^+, B^-, I^-, B^\leftrightarrow, I^\leftrightarrow\},
\]
where $B^+$ and $I^+$ denote boundary and interior births, $B^-$ and
$I^-$ denote boundary and interior deaths, and $B^\leftrightarrow$ and $I^\leftrightarrow$
denote boundary and interior moves. Let
$p_{B^+}$, $p_{I^+}$, $p_{B^-}$, $p_{I^-}$, $p_{B^\leftrightarrow}$, $p_{I^\leftrightarrow}$
be the corresponding baseline probabilities. At the current state
$\Delta$, unavailable proposal families are removed and the remaining
probabilities are renormalized. Specifically, $B^+$ and $I^+$ are always
available, $B^-$ and $B^\leftrightarrow$ are available only if
$|\mathcal V_B|>0$, and $I^-$ and $I^\leftrightarrow$ are available only if
$|\mathcal V_I|>0$. We denote the resulting state-dependent probability
of selecting proposal family $a\in\mathcal M$ by
\[
\rho_a(\Delta)
=
\frac{p_a}{\sum_{b\in\mathcal M(\Delta)}p_b},
\qquad a\in\mathcal M(\Delta),
\]
where $\mathcal M(\Delta)\subset\mathcal M$ is the set of proposal families available at
$\Delta$. Each proposal is detailed below. Figure~\ref{fig:BDM_extended} provides a graphical illustration.

\begin{itemize}
\item \textit{Boundary birth $(B^+)$.}
A new boundary vertex is generated uniformly with respect to
arclength measure on $\partial\Omega$. The proposal density is
$$
q(\Delta'\mid\Delta)
=
\rho_{B^+}(\Delta)\frac{1}{|\partial\Omega|}.
$$

\item \textit{Interior birth $(I^+)$.}
A new interior vertex is generated uniformly over $\Omega$. The
proposal density is
$$
q(\Delta'\mid\Delta)
=
\rho_{I^+}(\Delta)\frac{1}{|\Omega|}.
$$

\item \textit{Boundary death $(B^-)$.}
This proposal is available only when $|\mathcal V_B|>0$. One boundary
vertex is selected uniformly from $\mathcal V_B$ and removed. The
proposal density is
$$
q(\Delta'\mid\Delta)
=
\rho_{B^-}(\Delta)\frac{1}{|\mathcal V_B|}.
$$

\item \textit{Interior death $(I^-)$.}
This proposal is available only when $|\mathcal V_I|>0$. One interior
vertex is selected uniformly from $\mathcal V_I$ and removed. The
proposal density is
$$
q(\Delta'\mid\Delta)
=
\rho_{I^-}(\Delta)\frac{1}{|\mathcal V_I|}.
$$

\item \textit{Boundary move $(B^\leftrightarrow)$.}
This proposal is available only when $|\mathcal V_B|>0$. One boundary
vertex is selected uniformly from $\mathcal V_B$. Suppose the selected
vertex $\mathbf v\in\mathcal V_B$ lies on boundary segment $[\mathbf v_1,\mathbf v_2]$ and write
$
\mathbf v=(1-t)\mathbf v_1+t \mathbf v_2$ for $t\in(0,1)$.
We draw $\delta$ from a uniform distribution on the interval $[-h_B,h_B]$
and propose
$$
t'=t+\delta,
\qquad
\mathbf v'=(1-t')\mathbf v_1+t'\mathbf v_2,
$$
where $h_B>0$ is a tuning parameter. If $t'\notin(0,1)$, the proposal is
rejected. Otherwise, the selected vertex $\mathbf v$ is moved to $\mathbf v'$ and the CDT is
recomputed. With respect to arclength measure on the boundary segment,
the proposal density is
$$
q(\Delta'\mid\Delta)
=
\rho_{B^\leftrightarrow}(\Delta)
\frac{1}{|\mathcal V_B|}\times
\frac{1}{2h_B\|\mathbf v_1-\mathbf v_2\|_2}.
$$
For admissible boundary moves, this proposal is symmetric.

\item \textit{Interior move $(I^\leftrightarrow)$.}
This proposal is available only when $|\mathcal V_I|>0$. One interior
vertex $\mathbf v$ is selected uniformly from $\mathcal V_I$. We draw uniformly $\mathbf u$ on the ball $\{\mathbf z\in\mathbb R^2:\|\mathbf z\|_2\le r_I\}$
and propose $\mathbf v' = \mathbf v+\mathbf u$.
where $r_I>0$ is a tuning parameter. If $\mathbf v'\notin\Omega$, the proposal
is rejected. Otherwise, the selected vertex $\mathbf v$ is moved to $\mathbf v'$ and the CDT
is recomputed. The proposal density is
\[
q(\Delta'\mid\Delta)
=
\rho_{I^\leftrightarrow}(\Delta)
\frac{1}{|\mathcal V_I|}
\frac{1}{\pi r_I^2}.
\]
For admissible interior moves, this proposal is symmetric.
\end{itemize}

Each proposed triangulation is accepted with probability
\[
\alpha(\Delta,\Delta')
=
\min\!\left\{
1,\,
\frac{p(\mathbf y\mid\Delta',\nu,\lambda)}
     {p(\mathbf y\mid\Delta,\nu,\lambda)}
\frac{
a_B(V_B')a_I(V_I')
}{
a_B(V_B)a_I(V_I)
}
\frac{
|\partial\Omega|^{V_B}|\Omega|^{V_I}
}{
|\partial\Omega|^{V'_B}|\Omega|^{V'_I}
}
\frac{q(\Delta\mid\Delta')}{q(\Delta'\mid\Delta)}
\right\},
\]
where \(V_B=|\mathcal V_B|\), \(V_I=|\mathcal V_I|\),
\(V'_B=|\mathcal V'_B|\), and \(V'_I=|\mathcal V'_I|\), with the
convention \(0\log 0=0\). If the proposed triangulation is not admissible,
the proposal is rejected.

\section{Posterior Contraction Rates}
\label{sec:rate}

In this section, we establish posterior contraction rates for the proposed method. With respect to a given semimetric, a posterior contraction rate quantifies the speed at which the posterior distribution converges to the true parameter. In Section~\ref{sec:minimax}, we first show that for a Sobolev class with global smoothness over $\Omega$, the proposed procedure achieves the minimax rate up to a logarithmic factor. In Section~\ref{sec:spatialada}, we demonstrate that the proposed method exhibits ideal spatial adaptation \citep{donoho1994ideal} by attaining the near-oracle risk over all triangulations with weak shape regularity.

\subsection{Minimax Estimation under Global Smoothness}
\label{sec:minimax}

We first show that the proposed method achieves a near-minimax posterior contraction rate under global Sobolev smoothness over $\Omega$. We assume that the true function $f_0$ belongs to the Sobolev space $\mathcal W^{m,\infty}$ for some $m\in\mathbb N$, so that all weak derivatives of $f_0$ up to order $m$ have finite $L^\infty$-norms. A key step in establishing the contraction rate is to construct an approximant of $f_0$ for which the prior places sufficient mass on a Kullback--Leibler neighborhood of $f_0$. Classical spline approximation theory implies that, for shape-regular triangulations, the approximation error is of order $L_\Delta^m$, where $L_\Delta$ is the maximum edge length of $\Delta$ \citep{lai1998approximation}. Since our prior controls triangulation complexity through the number of vertices rather than through the maximum edge length, we need to characterize the approximation resolution in terms of the number of vertices. To this end, we show that there exists a well-behaved vertex set such that, for the CDT $\Delta$ generated from any neighboring vertex set, $L_\Delta$ is of the same order as $V_\Delta^{-1/2}$; see the supplementary material.

Using this optimal approximator, we establish the posterior contraction rate through the standard argument based on prior concentration and suitable tests \citep{ghosal2000convergence,ghosal2007convergence}. This argument requires a test function that is exponentially powerful with respect to the chosen metric. Although the Hellinger distance is well suited for constructing such tests \citep{lecam1973convergence,birge1983robust}, its use in regression settings often requires relatively strong boundedness conditions \citep{ghosal2007convergence}. To accommodate the unknown variance, we adopt the test function constructed by \citet{jeong2025l2}, which employs a metric that jointly measures the discrepancy in the mean function and the variance parameter in Gaussian regression. Combining this test function with the entropy bound for a suitably chosen sieve yields the posterior contraction rate under the prior specified in Section~\ref{sec:prior}. The proof is given in the supplementary material.

\begin{theorem}[Posterior contraction; Sobolev]
\label{thm:rate}
        Assume $f_0\in\mathcal W^{m,\infty}$ with $m\in\mathbb N$, and let $d\in\mathbb N$ and $r\in\mathbb N$ satisfy $d+1\ge m$ and $d\ge3r+2$. Suppose that the prior is specified for $\mathcal T_\Omega^\CD(\zeta_0,\ell_n)$ as in Section~\ref{sec:prior}, where \begin{align*}
            \zeta_0\le \frac{1}{2}\arcsin\!\left(\frac{1}{\sqrt{2}}\sin(\min\{\zeta_\Omega,\pi/3\})\right),\quad n^{-c}\lesssim \ell_n\lesssim n^{-1/4},
        \end{align*}
        with $c>1/4$, where $\zeta_\Omega$ is the minimum interior angle of the polygonal domain $\Omega$.
        Then, for every $M_n\rightarrow\infty$, the posterior distribution satisfies
    \begin{align*}
        \mathbb E_0\Pi\!\left
        \{\lVert f-f_0\rVert_n + \lvert \sigma^2-\sigma^2_0\rvert > M_n\bigg(\frac{\log n}{n}\bigg)^{m/(2m+2)} \, \Big| \, \mathbf y\right\}\rightarrow 0.
    \end{align*}
\end{theorem}

Theorem~\ref{thm:rate} shows that the posterior distribution contracts around the true parameter at the minimax rate for two-dimensional function estimation up to a logarithmic factor \citep{stone1982optimal}. The condition on $\zeta_0$ is mild because we can choose a sufficiently small positive minimum-angle threshold. The theoretical argument only needs $\zeta_0$ to be strictly positive, while the displayed upper bound ensures that the well-behaved CDT construction is contained in the support of the prior. The condition on $\ell_n$ requires the minimum admissible edge length to decrease at a suitable polynomial rate. It must decrease fast enough to allow the optimal approximating triangulation, but not so fast that the entropy bound becomes too large. Since $\ell_n$ is allowed to converge to zero at a sufficiently fast polynomial rate, this is merely a mild technical condition in practice.

\subsection{Ideal Spatial Adaptation}
\label{sec:spatialada}

Section~\ref{sec:minimax} establishes optimal performance in a worst-case sense when the underlying function $f_0$ is globally regular. However, such a global smoothness result does not capture the ability of the proposed method to adapt to inhomogeneous or spatially varying features of $f_0$. In this section, we show that the proposed method indeed achieves ideal spatial adaptation in the sense of \citet{donoho1994ideal}. More precisely, although the prior is supported on CDTs, the resulting posterior contraction rate is governed by an oracle benchmark defined over an arbitrary class of triangulations of $\Omega$. % This result demonstrates spatial adaptation, as the posterior can adapt to the spatially varying features of $f_0$ while controlling the complexity of the selected triangulation. 

Let $\mathfrak T_\Omega$ be a collection of triangulations of $\Omega$. For $f_0:\Omega\rightarrow \mathbb R$, the benchmark oracle empirical $L^2$-risk is defined by
\begin{align}
R_n(f_0; \mathfrak T_\Omega)
= \inf_{\Delta\in\mathfrak T_\Omega}\left\{\inf_{f\in\mathcal S_d^r(\Delta)}\lVert f-f_0\rVert_n^2
+ \frac{\sigma_0^2 J_\Delta}{n}\right\}.
\label{eqn:oraclerisk}
\end{align}
This quantity balances the approximation error of $f_0$ within the spline space $\mathcal S_d^r(\Delta)$ and the stochastic error associated with estimating the spline coefficients, with $J_\Delta$ serving as the dimension-based complexity penalty. 

Unlike the minimax benchmark in Section~\ref{sec:minimax}, the oracle risk in \eqref{eqn:oraclerisk} depends on the local and global features of the particular function $f_0$, and hence the benchmark reflects how well the triangulations in $\mathfrak T_\Omega$ can represent the spatial structure of $f_0$. In particular, the oracle may favor triangulations that allocate higher resolution to regions with localized features, while the penalty term $\sigma_0^2J_\Delta/n$ controls the total complexity of the resulting spline space.
The oracle risk $R_n(f_0; \mathfrak T_\Omega)$ therefore provides a natural benchmark for procedures whose model class is restricted to $\mathfrak T_\Omega$. Because the triangulation prior is supported on 
$\mathcal T_{\Omega}^\CD(\zeta_0,\ell_n)$, 
the proposed method should ideally attain a near-oracle rate over a class of triangulations that closely reflects this support.

We identify the class of triangulations for which the proposed method attains near-optimality.
For two finite point sets $\mathcal V,\mathcal V'\subset\mathbb R^2$ with $|\mathcal V|=|\mathcal V'|$, define
$$
d_{\text{match}}(\mathcal V,\mathcal V')
:=
\min_{\pi:\mathcal V\to \mathcal V'}
\max_{\mathbf v\in \mathcal V}
\|\mathbf v-\pi(\mathbf v)\|_2,
$$
where $\pi$ is a bijection.
For vertex sets $\mathcal V_B\subset \partial\Omega$ and $\mathcal V_I\subset\Omega$, define
\begin{align*}
\mathfrak V_\delta(\mathcal V_B,\mathcal V_I)=
\Big\{
(\mathcal V_B',\mathcal V_I') : 
& \ \mathcal V_B'\subset \partial\Omega, \
\mathcal V_I'\subset\Omega,\
|\mathcal V_B'|=|\mathcal V_B|,\
|\mathcal V_I'|=|\mathcal V_I|, \\
& \ d_{\mathrm{match}}(\mathcal V_B,\mathcal V_B')\le \delta,\
d_{\mathrm{match}}(\mathcal V_I,\mathcal V_I')\le \delta, \ \mathcal V_\Omega\cap\mathcal V_B'=\varnothing
\Big\}.
\end{align*}
Let $\widetilde{\mathcal T}_\Omega^{\CD}(\zeta,\ell,\delta) $ be defined as the collection of triangulations $\Delta_\Omega^\CD(\mathcal V_B,\mathcal V_I)\in\mathcal T_\Omega^{\CD}(\zeta,\ell)$ such that, 
for every pair of vertex sets 
$(\mathcal V_B',\mathcal V_I')\in \mathfrak V_\delta(\mathcal V_B,\mathcal V_I)$, the perturbed CDT $\Delta_\Omega^\CD(\mathcal V_B',\mathcal V_I')$ satisfies $\Delta_\Omega^\CD(\mathcal V_B',\mathcal V_I')\in \mathcal T_\Omega^{\CD}(\zeta,\ell)$ and has the same combinatorial structure as $\Delta_\Omega^\CD(\mathcal V_B,\mathcal V_I)$. Hence, $\widetilde{\mathcal T}_\Omega^{\CD}(\zeta,\ell,\delta)$ consists of triangulations such that every $\delta$-perturbation of their vertex sets still induces a CDT in $\mathcal T_\Omega^{\CD}(\zeta,\ell)$ with the same combinatorial structure. By definition,
$\widetilde{\mathcal T}_\Omega^{\CD}(\zeta,\ell,\delta)
\subset
\mathcal T_\Omega^{\CD}(\zeta,\ell)$, and we have
$
\widetilde{\mathcal T}_\Omega^{\CD}(\zeta,\ell,0)
=
\mathcal T_\Omega^{\CD}(\zeta,\ell)
$.
For small $\delta>0$, the robust core $\widetilde{\mathcal T}_\Omega^{\CD}(\zeta,\ell,\delta)$ may be viewed as a conservative inner approximation to $\mathcal T_\Omega^{\CD}(\zeta,\ell)$. It excludes triangulations lying too close to the boundary of the admissible class or to a change in CDT combinatorial structure, while retaining triangulations that are stable under small vertex perturbations.

The following theorem shows that the proposed method attains a near-oracle rate over this robust core. The proof is provided in the supplementary material.

\begin{theorem}[Posterior contraction; spatial adaptation]
\label{thm:oraclerate}
Assume that $\|f_0\|_{L^\infty}<\infty$, and let $d\in\mathbb N$ and $r\in\mathbb N$ satisfy
$d\ge 3r+2$. Suppose that the prior is specified as in
Section~\ref{sec:prior}, with support
$\mathcal T_\Omega^{\CD}(\zeta_0,\ell_n)$ for a fixed $\zeta_0>0$.
Assume that $\ell_n\ge n^{-A_\ell}$ and $\delta_n\ge n^{-A_\delta}$ for some constants $A_\ell>0$ and $A_\delta>0$.
For $B\ge 2\lVert f_0\rVert_{L^\infty}$, define
$$
r_n^2 
= \inf_{\Delta\in\widetilde{\mathcal T}_\Omega^{\CD}(\zeta_0,\ell_n,\delta_n)}\left\{\inf_{f\in\mathcal S_d^r(\Delta):\lVert f \rVert_{L^\infty}\le B}\lVert f-f_0\rVert_n^2
+ \frac{\sigma_0^2 J_\Delta\log n}{n}\right\}.
$$
Assume that $r_n\to0$ and $n r_n^2\to\infty$. Then, for every
$M_n\to\infty$,
$$
\mathbb E_0\Pi\!\left\{
\|f-f_0\|_n+\left|\sigma^2-\sigma_0^2\right|
>
M_n r_n
\,\Big|\,\mathbf y
\right\}
\to0 .
$$
\end{theorem}

The rate $r_n$ is slightly weaker than the oracle rate
$R_n^{1/2}(f_0; \mathcal T_{\Omega}^\CD(\zeta_0,\ell_n))$. First, the approximation term is restricted to splines satisfying the uniform bound $\lVert f\rVert_{L^\infty}\le B$. Second, the stochastic term contains an additional logarithmic factor. These two differences are mild because $f_0$ is assumed to be uniformly bounded and the logarithmic factor is negligible at the level of near-oracle rates. A more visible difference is that $r_n$ is defined over the perturbation-stable subcollection
$\widetilde{\mathcal T}_\Omega^{\CD}(\zeta_0,\ell_n,\delta_n)$ rather than over the entire CDT class
$\mathcal T_{\Omega}^\CD(\zeta_0,\ell_n)$. This stability restriction is mild because $\delta_n$ decreases polynomially. Hence the robustness condition is imposed only against increasingly small vertex perturbations and becomes weaker as $n$ grows.
Nevertheless, one might still worry that the proposed triangulation approach achieves near-optimality only over a restricted subset of the prior support. The following theorem shows that this is not the case. Up to constant losses in the shape-regularity parameters, the resulting rate is controlled by the near-oracle risk over arbitrary triangulations of $\Omega$ satisfying mild shape-regularity conditions. The proof is given in the supplementary material.

\begin{theorem}[Oracle risks over arbitrary triangulations]
Let $\mathcal T_\Omega^\ast (\zeta,\ell)$ be the collection of all triangulations of $\Omega$ with minimum angle $\zeta>0$ and minimum edge length $\ell>0$. There exists a constant $C$, depending only on $\zeta_0$, such that
    \begin{align*}
r_n^2 &\lesssim \inf_{\Delta\in\mathcal T_\Omega^\ast (C\zeta_0,C\ell_n)}\left\{\inf_{f\in\mathcal S_d^r(\Delta):\lVert f \rVert_{L^\infty}\le B}\lVert f-f_0\rVert_n^2
+ \frac{\sigma_0^2 J_\Delta\log n}{n}\right\}.
\end{align*}
\end{theorem}

\section{Simulation Study}
\label{sec:sim}

In this section, we conduct simulation studies to assess the empirical performance of BTS under several data-generating mechanisms. We consider two true regression functions $f_0$ on a horseshoe-shaped domain $\Omega$, as illustrated in the top-left panels in Figures~\ref{fig:compare_pred_ft} and \ref{fig:compare_pred_ex}.
The first function is globally smooth and has a comparable level of complexity across the domain, but its function values vary monotonically along the horseshoe strip. This setting is designed to examine whether the smoothing procedure respects the geometry of the domain, in particular whether the two arms of the horseshoe are treated as separated regions rather than being artificially smoothed across the gap.
The second function is also smooth, but exhibits spatially varying complexity, with relatively rapid local variation in some parts of the domain and smoother behavior elsewhere. This setting is used to evaluate whether a method can achieve spatial adaptation in an empirical sense.
The design points $\mathbf s_i$ were drawn uniformly from $\Omega$, and the responses were generated according to
the model in \eqref{eq:modelbig}.
We considered two true noise levels, $\sigma_0=0.1$ and $\sigma_0=0.5$, and two training sample sizes, $n=5{,}000$ and $n=20{,}000$.

We compare BTS with four competing approaches for nonparametric regression on irregular domains: SFS \citep{wood2008soap}, BPST \citep{yu2020estimation}, BAST \citep{luo2021bast}, and BORA-GP \citep{jin2024spatial}. BPST requires a pre-specified triangulation, and its performance may depend on this choice. We therefore consider two triangulations containing 101 and 242 triangles, denoted by BPST1 and BPST2, respectively. BAST requires specifying the number of weak learners, which affects both predictive performance and computational cost. We consider BAST with 10 and 30 weak learners, denoted by BAST1 and BAST2, respectively. The performance of BORA-GP is influenced by the number of neighbors. We consider BORA-GP with 10 and 20 neighbors, denoted by BORA-GP1 and BORA-GP2, respectively.

\begin{figure}[t!]
  \centering
  % First subfigure
  \begin{subfigure}{0.48\textwidth}
    \centering
    \includegraphics[width=\textwidth]{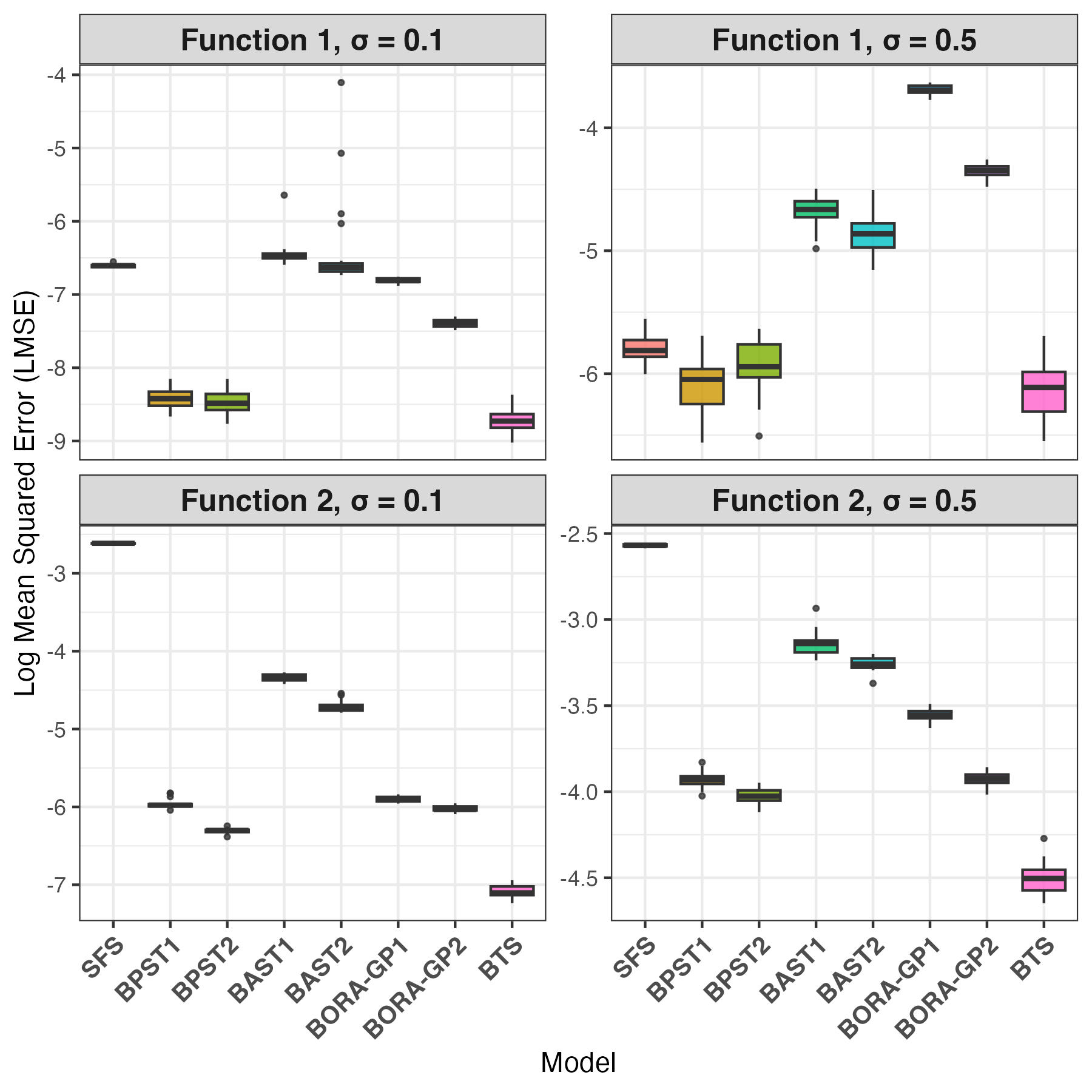}
    \label{fig:lmse_n5000}
    \caption{Estimation accuracy with $n=5{,}000$.}
  \end{subfigure}\hfill
  % Second subfigure
  \begin{subfigure}{0.48\textwidth}
    \centering
    \includegraphics[width=\textwidth]{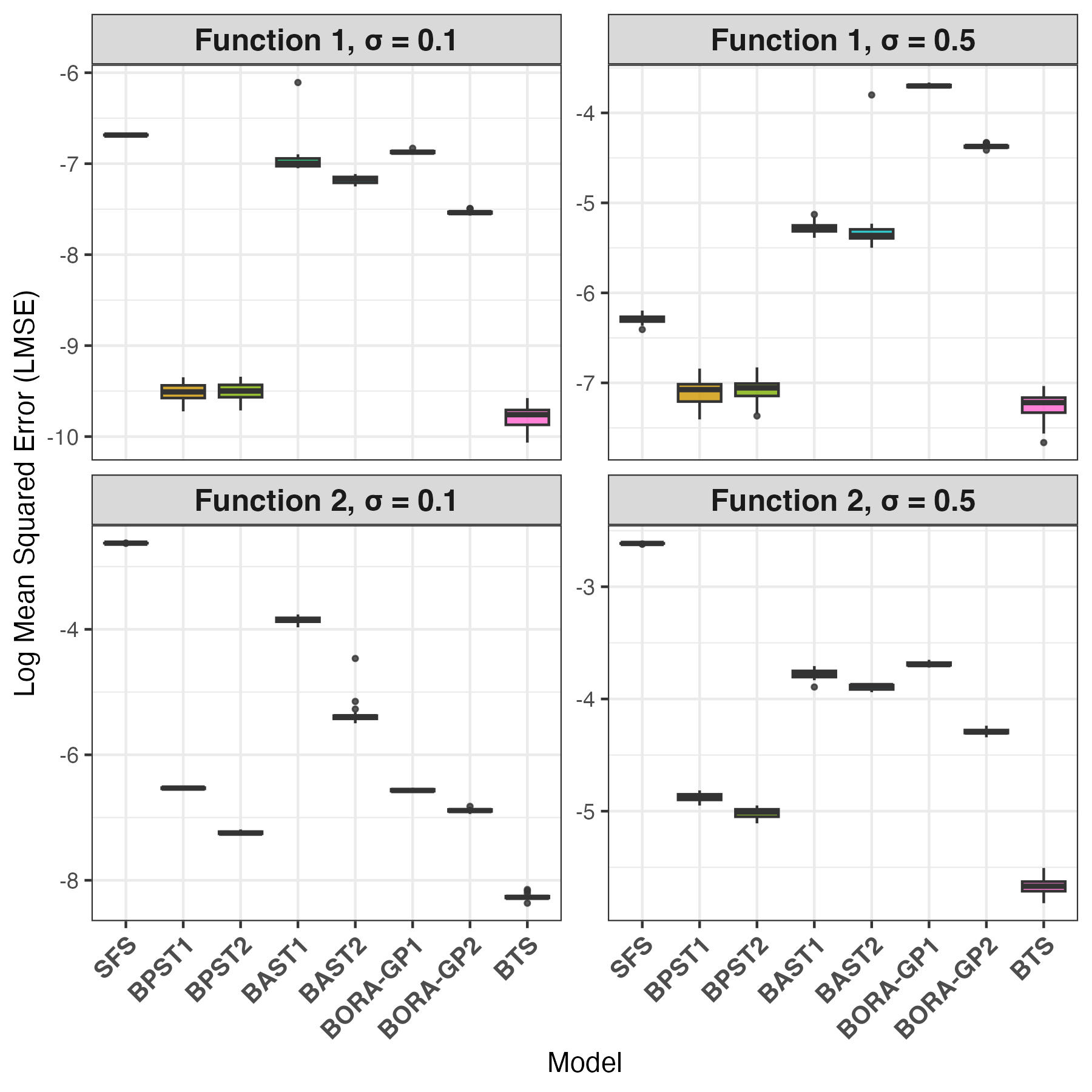}
    \label{fig:lmse_n2e4}
    \caption{Estimation accuracy with $n=20{,}000$.}
  \end{subfigure}
  \caption{Boxplots of the logarithm of MSE values across 25 replicated datasets.}
  \label{fig:lmse}
\end{figure}

For each simulation scenario, we generate 25 replicated datasets. In each replication, each method is fitted to the data, and a pointwise function estimate $\hat f$ is obtained. Performance is evaluated using the mean squared error (MSE), defined as $\|\hat f-f_0\|_{L^2}^2$.
Figure~\ref{fig:lmse} presents box plots of the logarithm of the MSE values across 25 replications. Overall, BTS consistently outperforms the competing approaches across all simulation settings. The performance gap is more pronounced for the target function with spatially varying complexity. This result suggests that BTS effectively captures local features of the target function in finite samples, as illustrated in Figure~\ref{fig:compare_pred}, and is consistent with the theoretical guarantee of spatial adaptation established in Section~\ref{sec:spatialada}. In contrast, the competing methods are less effective at adapting to local complexity. As shown in Figure~\ref{fig:compare_pred}, some methods tend to oversmooth locally complex regions, whereas others exhibit overly variable estimates, possibly due to the prespecification of the estimation resolution or the attempt to accommodate local features through fixed tuning parameters. BTS avoids these issues by adapting the triangulation structure to the data.

\begin{figure}[t!]
  \centering
  % First subfigure
  \begin{subfigure}{0.48\textwidth}
    \centering
    \includegraphics[width=\textwidth]{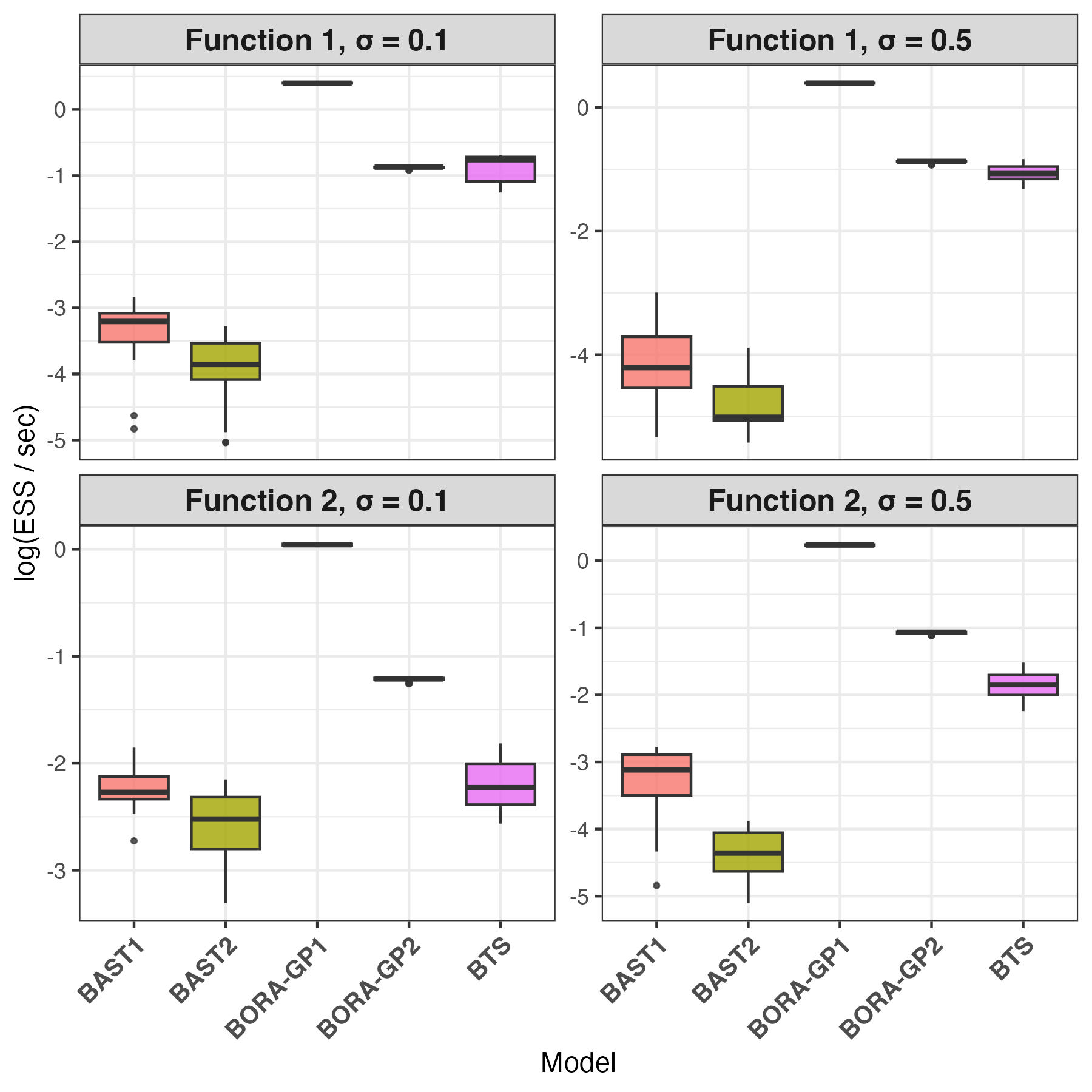}
    \label{fig:logess_n5000}
    \caption{Sampling efficiency with $n=5{,}000$.}
  \end{subfigure}\hfill
  % Second subfigure
  \begin{subfigure}{0.48\textwidth}
    \centering
    \includegraphics[width=\textwidth]{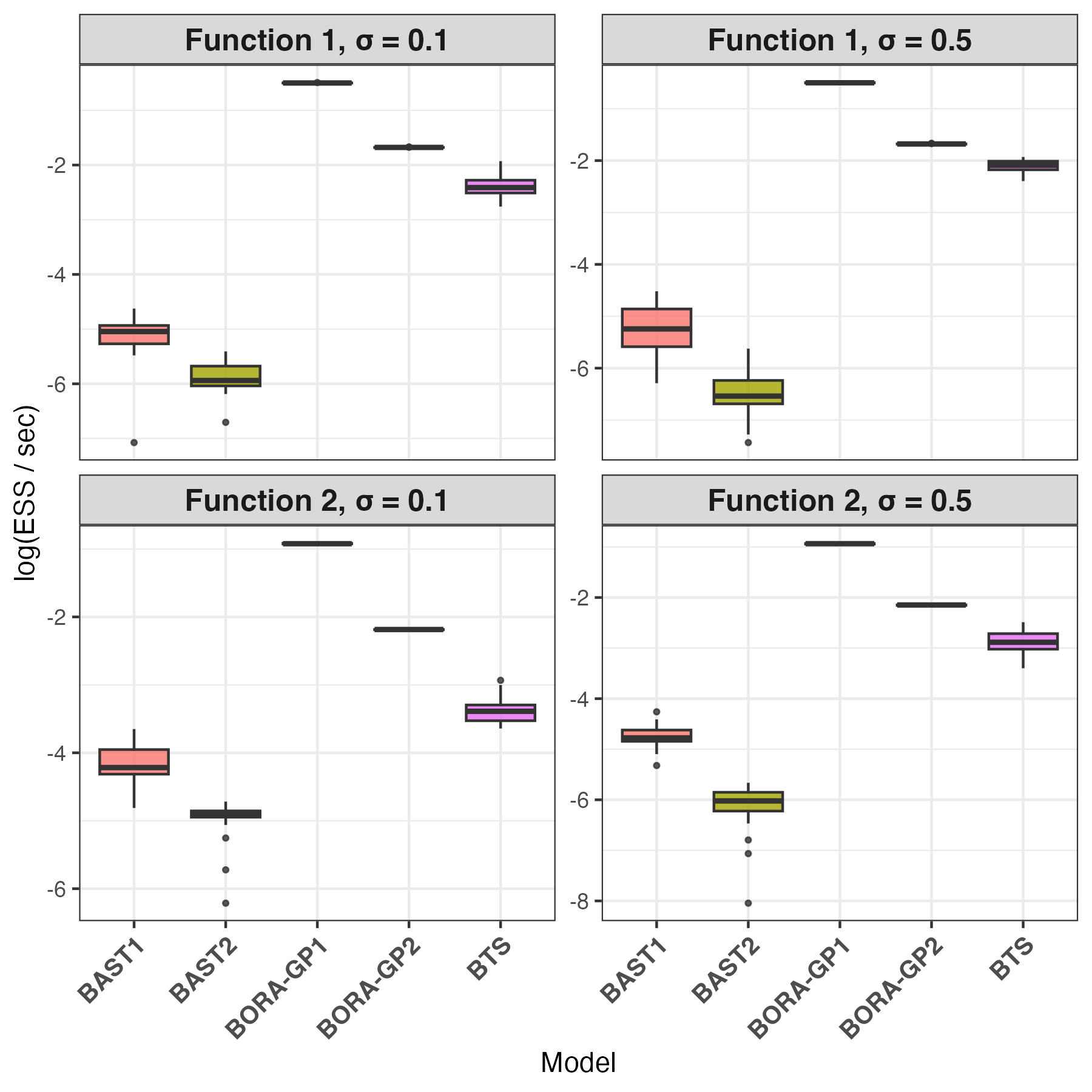}
    \label{fig:logess_n2e4}
    \caption{Sampling efficiency with $n=20{,}000$.}
  \end{subfigure}
  \caption{Boxplots of the logarithm of ESS per second for the MCMC-based Bayesian methods across 25 replicated datasets.}
  \label{fig:logess}
\end{figure}

We also compare the sampling efficiency of the MCMC-based Bayesian methods, including BTS, BAST, and BORA-GP. SFS and BPST are excluded from this comparison because they are non-adaptive frequentist methods and are therefore not directly comparable in terms of MCMC sampling efficiency. 
For each method, posterior function samples are evaluated on a common grid over the domain, and the pointwise effective sample size (ESS) of the resulting function values is computed at each grid point. The ESS per second is then averaged over the grid points. Figure~\ref{fig:logess} reports this average ESS per second across 25 replicated datasets. Although BORA-GP tends to have slightly higher ESS per second, BTS remains computationally competitive while achieving substantially better predictive accuracy, as shown in Figure~\ref{fig:lmse}. By contrast, BAST is substantially less efficient than the other MCMC-based methods and also lags behind in predictive performance.

%%%%%%%%%%%%%%%%%%%%%%%%%%%%%%%%%%%%%%%%%%%%%%%%%%%%%%
\section{Application to the Sea of Azov Data}
\label{sec:realdata}

In this section, we apply BTS to analyze chlorophyll-\textit{a} concentrations
in the Sea of Azov and the Black Sea. Monitoring chlorophyll concentration is
important because it serves as a proxy for phytoplankton biomass, which plays a
central role in marine ecosystems and the global climate system. However,
chlorophyll concentration is influenced by multiple environmental factors,
including water flow, river discharge, and temperature. Consequently, chlorophyll concentrations can exhibit pronounced local spatial
variability, particularly near regions affected by strong water exchange or
river inflow.
We analyze Level-3 monthly data from NASA's Aqua-MODIS satellite, covering the
period from May 17 to June 17, 2013, at a spatial resolution of 4 km. 
The dataset consists of $3{,}298$ observations, each with corresponding latitude, longitude, and chlorophyll-\textit{a} concentration level.  The dataset
is available from NASA Ocean Color\footnote{\url{https://oceancolor.gsfc.nasa.gov/resources/docs/tutorials/notebooks/modis-explore-l3/}}.

\begin{figure}[t!]
    \centering
    \includegraphics[width=0.7\linewidth]{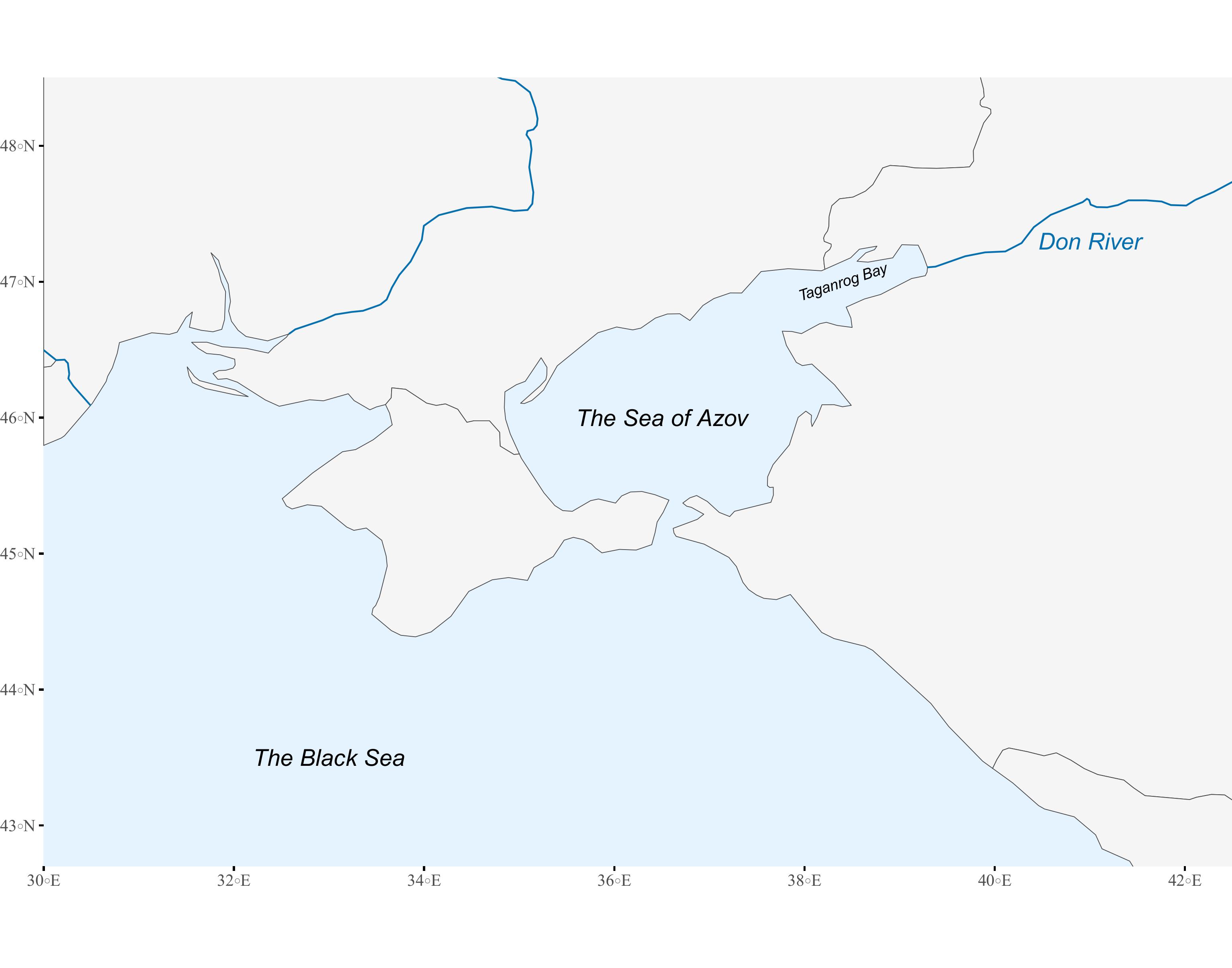}
    \caption{Map of the Sea of Azov and surrounding regions.}
    \label{fig:azov_map}
\end{figure}

\begin{figure}[t!]
  \centering
  % First subfigure
  \begin{subfigure}{0.49\textwidth}
    \centering
    \includegraphics[width=\textwidth]{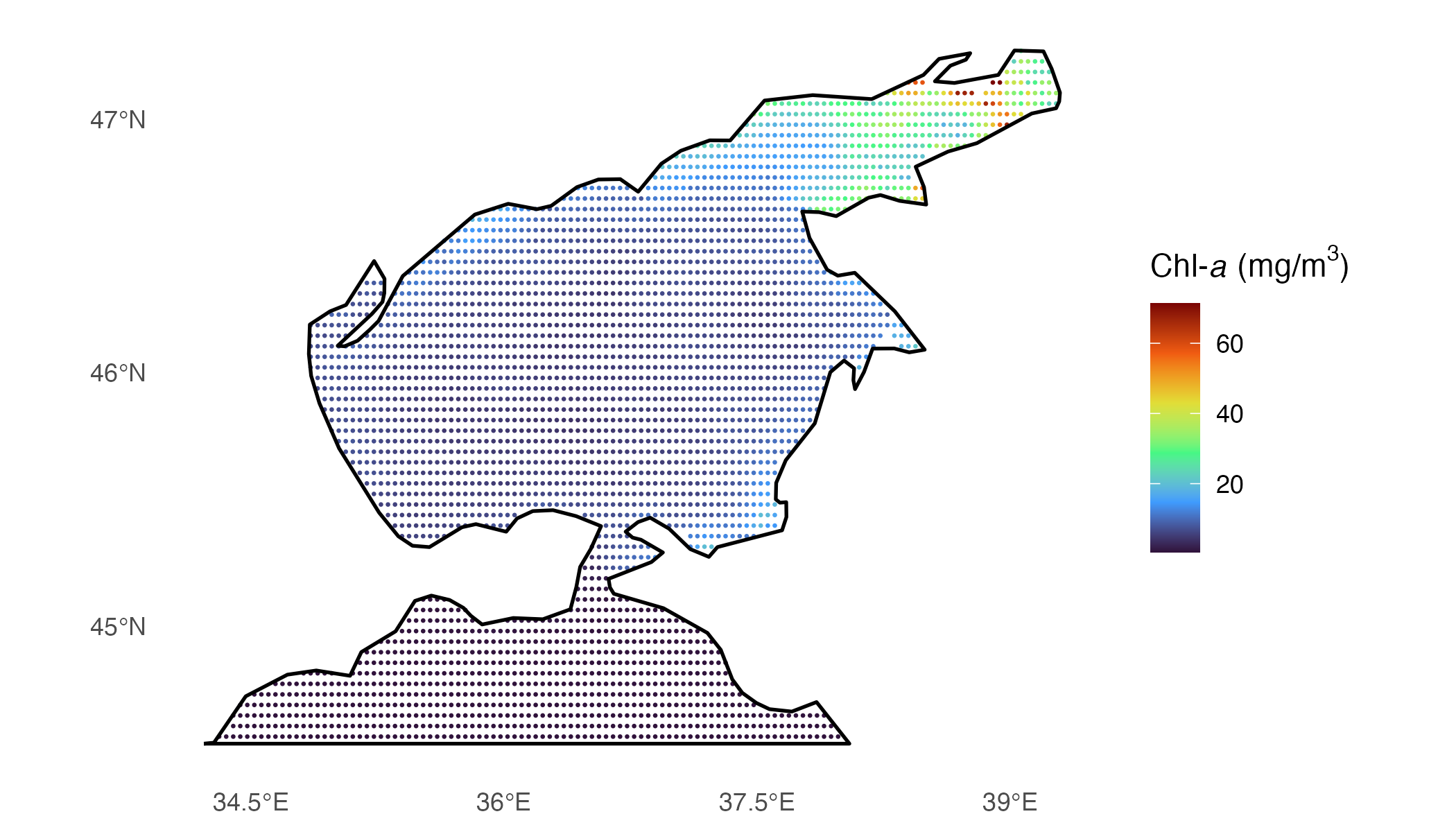}
    \caption{Observed chlorophyll concentration.}
    \label{fig:azov_train2}
  \end{subfigure}
  % Second subfigure
  \begin{subfigure}{0.49\textwidth}
    \centering
    \includegraphics[width=\textwidth]{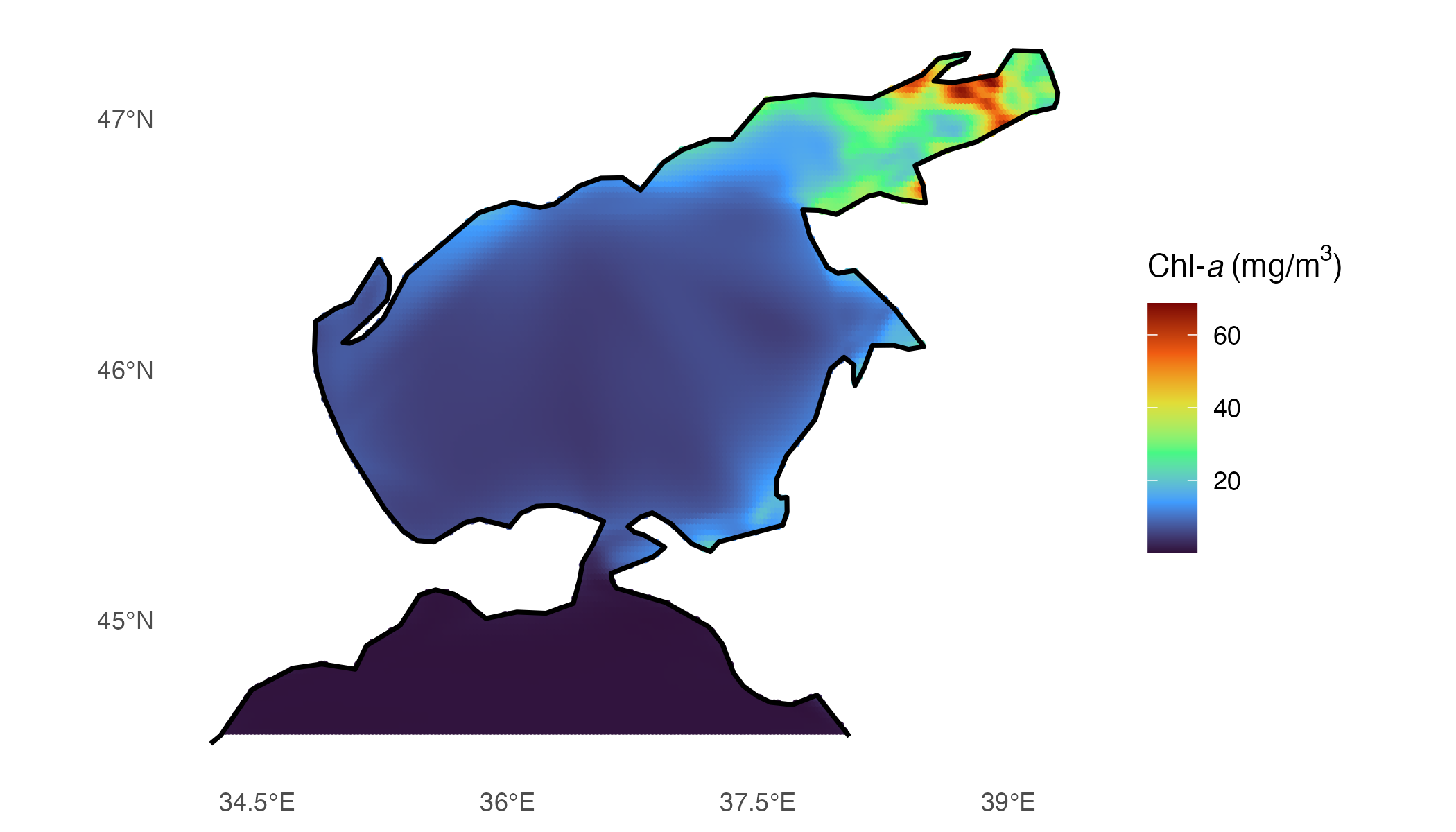}
    \caption{Pointwise posterior mean estimate using BTS.}
    \label{fig:azov_pred2}
  \end{subfigure}
  \caption{Observed chlorophyll concentration in the training dataset and function estimate obtained by BTS.}
  \label{fig:azov_traintest2}
\end{figure}

As illustrated in Figure~\ref{fig:azov_map}, the Sea of Azov has a
hydrographically distinctive setting. It is connected to the Black Sea only
through a narrow strait known as the Kerch Strait, so seawater exchange between the two basins is
strongly constrained by this passage. In addition, the Sea of Azov receives freshwater input from several rivers and
tributaries. The largest contribution comes from the Don River, which discharges
into Taganrog Bay. In particular, this strong freshwater input near Taganrog Bay
can induce pronounced local variation in chlorophyll concentration. Near this
major river mouth, environmental conditions such as nutrient availability,
turbidity, salinity, and vertical mixing may change sharply over short spatial
scales.
Consequently, chlorophyll concentrations can exhibit substantial local
variability in these areas. The observed chlorophyll concentration shown in
Figure~\ref{fig:azov_train2} 
is consistent with this expectation, exhibiting
spatially heterogeneous patterns over the study region. These features make the
dataset particularly suitable for illustrating the advantages of BTS, which can
incorporate complex domain boundaries while allowing spatially adaptive
smoothing.

To avoid the adverse effects of using an overly complex domain polygon with too many corner vertices, we constructed a parsimonious polygonal domain for the real-data analysis. This was done by buffering the original coastline and then simplifying the boundary using the Visvalingam--Whyatt algorithm \citep{visvalingam1993line}. This yielded a robust polygonal approximation that covers the study region while retaining the main geometric features of the coastline. BTS was applied with the prior distribution specified in Section~\ref{sec:prior}. The pointwise posterior mean prediction is shown in Figure~\ref{fig:azov_traintest2}. The result shows that the Sea of Azov has consistently higher chlorophyll concentration than the Black Sea, with pronounced local fluctuations near Taganrog Bay. Overall, BTS captures both the broad spatial contrast between the two seas and the localized variability induced by the complex coastal and inflow structures.

\section{Discussion}
\label{sec:disc}

We proposed BTS for spatially adaptive
nonparametric regression on irregular domains. By placing a prior on both the
number and locations of triangulation vertices, the proposed method can control
global model complexity while locally refining the triangulation in regions
where the target function exhibits more complex spatial features. The use of
CDTs allows the procedure to respect domain
boundaries and avoid artificial smoothing across them.

The theoretical results show that BTS achieves near-minimax rate adaptation
under global Sobolev smoothness and also attains a near-oracle rate for
spatially inhomogeneous functions. The latter guarantee is particularly
important because the oracle benchmark is not restricted to the CDT class, but
is controlled by arbitrary shape-regular triangulations. The simulation studies
and real-data analysis support these theoretical findings, showing that BTS can
capture both large-scale spatial patterns and localized variation on complex
domains.

Several extensions are possible. Although we focused on Gaussian regression,
the proposed triangulation-based spline representation can be incorporated into
generalized regression models for non-Gaussian responses. It would also be
interesting to develop more scalable computational schemes for very large
datasets and to extend the framework to spatio-temporal settings where the
surface, the triangulation, or both may evolve over time.

\bibliographystyle{apalike}
\bibliography{ref2}

@article{stone1982optimal,
  title={Optimal global rates of convergence for nonparametric regression},
  author={Stone, Charles J},
  journal={The Annals of Statistics},
  pages={1040--1053},
  year={1982},
  publisher={JSTOR}
}

@incollection{birge1983robust,
  title={Robust testing for independent non identically distributed variables and Markov chains},
  author={Birg{\'e}, Lucien},
  booktitle={Specifying Statistical Models: From Parametric to Non-Parametric, Using Bayesian or Non-Bayesian Approaches},
  pages={134--162},
  year={1983},
  publisher={Springer}
}

@article{lecam1973convergence,
  title={Convergence of estimates under dimensionality restrictions},
  author={Le Cam, Lucien},
  journal={The Annals of Statistics},
  pages={38--53},
  year={1973},
  publisher={JSTOR}
}

@article{chipman2010bart,
  title={{BART}: Bayesian additive regression trees},
  author={Chipman, Hugh A and George, Edward I and McCulloch, Robert E},
  year={2010}
}

@article{zhou2001spatially,
  title={Spatially adaptive regression splines and accurate knot selection schemes},
  author={Zhou, Shanggang and Shen, Xiaotong},
  journal={Journal of the American Statistical Association},
  volume={96},
  number={453},
  pages={247--259},
  year={2001},
  publisher={Taylor \& Francis}
}

@article{miyata2003adaptive,
  title={Adaptive free-knot splines},
  author={Miyata, Satoshi and Shen, Xiaotong},
  journal={Journal of Computational and Graphical Statistics},
  volume={12},
  number={1},
  pages={197--213},
  year={2003},
  publisher={Taylor \& Francis}
}

@article{birge2001alternative,
  title={An alternative point of view on {L}epski's method},
  author={Birg{\'e}, Lucien},
  journal={Lecture Notes-Monograph Series},
  pages={113--133},
  year={2001},
  publisher={JSTOR}
}

@article{donoho1994ideal,
  title={Ideal spatial adaptation by wavelet shrinkage},
  author={Donoho, David L and Johnstone, Iain M},
  journal={Biometrika},
  volume={81},
  number={3},
  pages={425--455},
  year={1994},
  publisher={Oxford University Press}
}

@article{shen2015adaptive,
  title={Adaptive {B}ayesian procedures using random series priors},
  author={Shen, Weining and Ghosal, Subhashis},
  journal={Scandinavian Journal of Statistics},
  volume={42},
  number={4},
  pages={1194--1213},
  year={2015},
  publisher={Wiley Online Library}
}

@article{arbel2013bayesian,
  title={Bayesian optimal adaptive estimation using a sieve prior},
  author={Arbel, Julyan and Gayraud, Ghislaine and Rousseau, Judith},
  journal={Scandinavian Journal of Statistics},
  volume={40},
  number={3},
  pages={549--570},
  year={2013},
  publisher={Wiley Online Library}
}

@article{van2009adaptive,
  title={Adaptive {B}ayesian estimation using a {G}aussian random field with inverse gamma bandwidth},
  author={van der Vaart, Aad W and van Zanten, J Harry},
  year={2009}
}

@article{belitser2003adaptive,
  title={Adaptive {B}ayesian inference on the mean of an infinite-dimensional normal distribution},
  author={Belitser, Eduard and Ghosal, Subhashis},
  journal={The Annals of Statistics},
  volume={31},
  number={2},
  pages={536--559},
  year={2003},
  publisher={Institute of Mathematical Statistics}
}

@incollection{birge1997model,
  title={From model selection to adaptive estimation},
  author={Birg{\'e}, Lucien and Massart, Pascal},
  booktitle={Festschrift for Lucien Le Cam},
  pages={55--87},
  year={1997},
  publisher={Springer},
  editor={Pollard, David and Torgersen, Erik and Yang, Grace},
  address={New York}
}

@article{donoho1995adapting,
  title={Adapting to unknown smoothness via wavelet shrinkage},
  author={Donoho, David L and Johnstone, Iain M},
  journal={Journal of the American Statistical Association},
  volume={90},
  number={432},
  pages={1200--1224},
  year={1995},
  publisher={Taylor \& Francis}
}

@article{lepskii1991problem,
  title={On a problem of adaptive estimation in {G}aussian white noise},
  author={Lepskii, OV},
  journal={Theory of Probability \& Its Applications},
  volume={35},
  number={3},
  pages={454--466},
  year={1991},
  publisher={SIAM}
}

@article{bartlett1957comment,
    title={A comment on {D}. {V}. {L}indley's statistical paradox},
	author={M. S. Bartlett},
	journal={Biometrika},
	volume={44},
	number={3--4},
	pages={533–534},
	year={1957}
}

@article{moreno1998intrinsic,
  title={An intrinsic limiting procedure for model selection and hypotheses testing},
  author={Moreno, El{\'\i}as and Bertolino, Francesco and Racugno, Walter},
  journal={Journal of the American Statistical Association},
  volume={93},
  number={444},
  pages={1451--1460},
  year={1998},
  publisher={Taylor \& Francis}
}

@article{smith1996nonparametric,
  title={Nonparametric regression using {B}ayesian variable selection},
  author={Smith, Michael and Kohn, Robert},
  journal={Journal of Econometrics},
  volume={75},
  number={2},
  pages={317--343},
  year={1996},
  publisher={Elsevier}
}

@inproceedings{wang1987optimal,
  title={An optimal algorithm for constructing the {D}elaunay triangulation of a set of line segments},
  author={Wang, C and Schubert, Len},
  booktitle={Proceedings of the third annual symposium on Computational geometry},
  pages={223--232},
  year={1987}
}

@inproceedings{chew1987constrained,
  title={Constrained {D}elaunay triangulations},
  author={Chew, L Paul},
  booktitle={Proceedings of the Third Annual Symposium on Computational Geometry},
  pages={215--222},
  year={1987}
}

@book{cheng2013delaunay,
  title={Delaunay Mesh Generation},
  author={Cheng, Siu-Wing and Dey, Tamal Krishna and Shewchuk, Jonathan and Sahni, Sartaj},
  year={2013},
  publisher={CRC Press Boca Raton}
}

@article{yu2020estimation,
  title={Estimation and inference for generalized geoadditive models},
  author={Yu, Shan and Wang, Guannan and Wang, Li and Liu, Chenhui and Yang, Lijian},
  journal={Journal of the American Statistical Association},
  year={2020},
  publisher={Taylor \& Francis}
}

@article{lai2013bpst,
    title={Bivariate penalized splines for regression},
    author={Lai, Ming-Jun and Wang, Li},
    journal={Statistica Sinica},
    volume = {23},
    number = {3},
    pages={1399--1417},
    year={2013},
    publisher={JSTOR}
}

@article{lim2023synergizing,
  title={Penalty-Induced Basis Exploration for {B}ayesian Splines},
  author={Lim, Sunwoo and Pyeon, Sihyeon  and Jeong, Seonghyun},
  journal={arXiv preprint arXiv:2311.13481},
  year={2023}
}

@article{delaunay1934sphere,
  author    = {Boris Delaunay},
  title     = {Sur la sph{\`e}re vide. {\`A} la m{\'e}moire de Georges Vorono{\"\i}},
  journal   = {Izvestia Akademii Nauk SSSR, Otdelenie Matematicheskikh i Estestvennykh Nauk},
  volume    = {7},
  number    = {6},
  pages     = {793--800},
  year      = {1934},
  publisher = {USSR Academy of Sciences}
}

@article{wang2020efficient,
  title={Efficient estimation of partially linear models for data on complicated domains by bivariate penalized splines over triangulations},
  author={Wang, Li and Wang, Guannan and Lai, Ming-Jun and Gao, Lei},
  journal={Statistica Sinica},
  volume={30},
  number={1},
  pages={347--369},
  year={2020},
  publisher={JSTOR}
}

@article{wood2008soap,
  title={Soap film smoothing},
  author={Wood, Simon N and Bravington, Mark V and Hedley, Sharon L},
  journal={Journal of the Royal Statistical Society Series B: Statistical Methodology},
  volume={70},
  number={5},
  pages={931--955},
  year={2008},
  publisher={Oxford University Press}
}

@article{jin2024spatial,
  title={Spatial predictions on physically constrained domains: applications to Arctic sea salinity data},
  author={Jin, Bora and Herring, Amy H and Dunson, David},
  journal={The Annals of Applied Statistics},
  volume={18},
  number={2},
  pages={1596--1617},
  year={2024},
  publisher={Institute of Mathematical Statistics}
}

@article{ramsay2002spline,
  title={Spline smoothing over difficult regions},
  author={Ramsay, Tim},
  journal={Journal of the Royal Statistical Society Series B: Statistical Methodology},
  volume={64},
  number={2},
  pages={307--319},
  year={2002},
  publisher={Oxford University Press}
}

@article{niu2019intrinsic,
  title={Intrinsic {G}aussian processes on complex constrained domains},
  author={Niu, Mu and Cheung, Pokman and Lin, Lizhen and Dai, Zhenwen and Lawrence, Neil and Dunson, David},
  journal={Journal of the Royal Statistical Society Series B: Statistical Methodology},
  volume={81},
  number={3},
  pages={603--627},
  year={2019},
  publisher={Oxford University Press}
}

@article{denison1998automatic,
  title={Automatic {B}ayesian curve fitting},
  author={Denison, DGT and Mallick, BK and Smith, AFM},
  journal={Journal of the Royal Statistical Society: Series B (Statistical Methodology)},
  volume={60},
  number={2},
  pages={333--350},
  year={1998},
  publisher={Wiley Online Library}
}

@article{dimatteo2001bayesian,
  title={Bayesian curve-fitting with free-knot splines},
  author={DiMatteo, Ilaria and Genovese, Christopher R and Kass, Robert E},
  journal={Biometrika},
  volume={88},
  number={4},
  pages={1055--1071},
  year={2001},
  publisher={Oxford University Press}
}

@article{luo2021bast,
  title={{BAST}: Bayesian additive regression spanning trees for complex constrained domain},
  author={Luo, Zhao Tang and Sang, Huiyan and Mallick, Bani},
  journal={Advances in Neural Information Processing Systems},
  volume={34},
  pages={90--102},
  year={2021}
}

@article{ghosal2000convergence,
    title={Convergence rates of posterior distributions},
    author={Ghosal, Subhashis and Ghosh, Jayanta K and van der Vaart, Aad W},
    journal={Annals of Statistics}, 
    volume = {28},
    number = {2},
    pages={500--531},
    year={2000},
    publisher={JSTOR}
}

@article{ghosal2007convergence,
    title = {Convergence rates of posterior distributions for noniid observations},
    author = {Ghosal, Subhashis and van der Vaart, Aad},
    journal = {The Annals of Statistics},
    volume = {35},
    number = {1},
    pages = {192 -- 223},
    year = {2007},
    publisher = {Institute of Mathematical Statistics}
}

@article{jeong2025l2,
  title={L2-norm posterior contraction in {G}aussian models with unknown variance},
  author={Jeong, Seonghyun},
  journal={Statistics \& Probability Letters},
  volume={226},
  pages={110495},
  year={2025},
  publisher={Elsevier}
}

@article{lai1998approximation,
  title={On the approximation power of bivariate splines},
  author={Lai, Ming-Jun and Schumaker, Larry L},
  journal={Advances in Computational Mathematics},
  volume={9},
  pages={251--279},
  year={1998},
  publisher={Springer}
}

@book{lai2007spline,
  title={Spline Functions on Triangulations},
  author={Lai, Ming-Jun and Schumaker, Larry L},
  number={110},
  year={2007},
  publisher={Cambridge University Press}
}

@article{visvalingam1993line,
  title={Line generalisation by repeated elimination of points},
  author={Visvalingam, Maheswari and Whyatt, James D.},
  journal={The Cartographic Journal},
  volume={30},
  number={1},
  pages={46--51},
  year={1993},
  publisher={Maney Publishing},
  doi={10.1179/000870493786962263}
}

\end{document}